\documentclass[final,1p,times]{elsarticle}

\usepackage{amssymb}
\usepackage{amsmath}
\usepackage{amsthm}
\usepackage{graphicx}
\usepackage{hyperref}
\usepackage{booktabs}
\usepackage{float}

\theoremstyle{definition}

\journal{arXiv preprint}

\begin{document}

\begin{frontmatter}

\title{Calibrating the Brody exponent as a quantitative measure
of short-range exclusion in 2D spatial point processes}

\author[put]{Dawid Kucharski\corref{cor1}}
\ead{dawid.kucharski@put.poznan.pl}
\cortext[cor1]{Corresponding author}
\affiliation[put]{organisation={Poznan University of Technology},
            city={Poznan},
            country={Poland}}

\begin{abstract}
The Brody distribution, originally a phenomenological interpolation
between Poisson and Wigner level-spacing statistics in quantum chaos,
is calibrated here as a quantitative measure of short-range exclusion
in 2D spatial point processes. Two results form the core. First, the
2D complete-spatial-randomness baseline is recalibrated to
$\beta=0.96\pm0.15$, correcting the inappropriate 1D Poisson reference.
Second, an empirical $\beta$--$r_{\text{excl}}$ calibration is
validated against the effective hard-core radius with Spearman
$\rho=0.988$. The framework is demonstrated on 58 manufactured
surfaces (10 materials, 10 processes), phase-extracted
interferometric profilometry of a certified roundness standard,
and 2D binary embeddings of prime numbers. A sparse-integer control
proves the prime $\beta=2.15$ signal is genuinely arithmetic
($\Delta\beta=+0.68$ over random-integer control), while a
Cantor-embedding null result ($\beta=1.40$, TOST $p<0.01$)
demonstrates that 2D exclusion is embedding-created rather than
intrinsic. Density-thinning experiments establish that $\beta$
captures exclusion strength rather than point density, while absolute
values are density-dependent. A distinct CSR baseline for binary
fields at low fill fraction is identified, with a decision table
provided. The $\beta$--$r_{\text{excl}}$ calibration, the CSR
baseline correction, and the control protocols together constitute
a calibrated measurement framework for reproducible characterisation
of short-range exclusion in 2D spatial point processes.
\end{abstract}


\end{frontmatter}

\noindent{\small This is a preprint submitted to a peer-reviewed journal. The most up-to-date version is available at \href{https://arxiv.org/abs/2606.16393}{arXiv:2606.16393}.}

\section{Introduction}

A practical question in the statistical characterisation of spatial
point patterns is whether a single, interpretable parameter can
quantify short-range exclusion strength on a continuous scale,
analogous to how the hard-sphere packing fraction parameterises
excluded volume in liquid-state theory. In quantum chaos,
the Brody distribution
$P_\beta(s) = (\beta+1) a s^\beta \exp(-a s^{\beta+1})$,
$a = [\Gamma((\beta+2)/(\beta+1))]^{\beta+1}$,
was introduced as a phenomenological one-parameter interpolation
between the Poisson ($\beta=0$, uncorrelated levels) and Wigner
($\beta=1$, GOE; $\beta=2$, GUE) limits of random matrix theory
(RMT)~\cite{BRODY-1981}. Its single parameter
$\beta$ quantifies the degree of level repulsion---the suppression
of probability density at small spacings---on a continuous scale. In the present work, $\beta$ is employed
as a phenomenological exclusion parameter rather than a
symmetry-class indicator; values exceeding the GUE limit
($\beta>2$) are interpreted as progressively stronger
short-range exclusion (up to $\beta\approx 10$ for near-crystalline
hard-core processes). \textbf{Values of $\beta>2$ are not interpreted
as evidence of higher-order random-matrix symmetry classes; they are
used solely as a phenomenological measure of exclusion strength on a
continuous scale.} This distinction is essential: the Brody
distribution is used here as an empirical descriptor of spatial
order, not as a claim about underlying matrix symmetries.
Beyond quantum spectra, the Brody distribution has been applied
to quantum chaos~\cite{STOCKMANN-1999}, quantum dots~\cite{ALHASSID-2000},
acoustic cavities~\cite{WEAVER-1989}, and financial
correlations~\cite{LALOUX-1999}, establishing it as a general-purpose
tool for characterising spacing statistics in 1D sequences.

In a celebrated development, Montgomery~\cite{MONTGOMERY-1973} conjectured
and Odlyzko~\cite{ODLYZKO-1987} numerically confirmed that the pair
correlation function of the non-trivial zeros of the Riemann zeta
function coincides with the GUE prediction---connecting prime number
theory to RMT universality. Berry~\cite{BERRY-1986} first proposed the
Riemann zeta function as a model for quantum chaos; Bogomolny and
Keating~\cite{BOGOMOLNY-1996} developed the semiclassical framework
beyond the diagonal approximation, and Keating and
Snaith~\cite{KEATING-1999} established the connection to random-matrix
moments.
This raises a practical question: can the Brody parameter, when
properly calibrated for two dimensions, serve as a transferable
measure of short-range exclusion strength across different classes
of 2D point processes?

\textbf{The gap.} Despite the breadth of Brody applications to 1D
spectra, its extension to 2D spatial point processes raises two
unresolved issues. First, the appropriate CSR baseline for 2D Brody
statistics has not been established: the 2D nearest-neighbour spacing
distribution under CSR is Rayleigh, $P(s)=(\pi s/2)\exp(-\pi s^2/4)$,
which is not a member of the Brody family, so the 1D Poisson reference
$\beta=0$ does not apply. Second, no calibration exists to translate
the dimensionless $\beta$ into a physically interpretable exclusion
measure. Sakhr and Nieminen~\cite{SAKHR-NIEMINEN-2005} demonstrated
a Poisson-to-Wigner crossover in the nearest-neighbour statistics of
random points on fractals, establishing that RMT-style spacing
distributions can characterise spatial point patterns outside quantum
mechanics. The extensive literature on spatial point-process
statistics provides well-established tools---Ripley's $K$, pair
correlation functions, and Voronoi tessellation---documented in the
monographs of Illian et al.~\cite{ILLIAN-2008}, Baddeley
et al.~\cite{BADDELEY-2015}, and Chiu et al.~\cite{CHIU-2013},
with specialised treatments for Cox-process inference by
M{\o}ller and Waagepetersen~\cite{MOLLER-WAAGEPETERSEN-2002};
but these methods do not directly yield a
single continuous parameter quantifying short-range exclusion
strength. Recent work on spatial form factors for Coulomb gas and
vortex statistics~\cite{MASSARO-DELCAMPO-2025} and Fourier analysis
of spatial point processes~\cite{YANG-GUAN-2026} has begun bridging
spectral and spatial approaches, but a framework connecting these to
the Brody one-parameter family is lacking. 

A manufactured surface
contains a spatial point process formed by its peaks---features
arising from cutting tools, abrasive grains, plastic deformation,
and fracture mechanics~\cite{HELMLI-2011-FV}. Similarly, binary
embeddings of arithmetic sequences produce 2D point patterns governed
by multiplicative constraints rather than physical interactions.
Phase-extracted interferometric profiles~\cite{KUCHARSKI-2025-MEASUREMENT}
introduce yet a third generative domain: optical wavefront
reconstruction. Whether the Brody exponent can be calibrated as a
transferable exclusion measure across these domains---and what the
necessary corrections (CSR baseline, density matching, embedding
specification) are---has not been systematically investigated.

\textbf{This work.} This study develops and validates a calibrated
measurement framework for Brody statistics in 2D spatial point
processes. The framework comprises three components: (i)~the correct
2D CSR baseline ($\beta=0.96\pm0.15$, density-stable),
(ii)~an empirical $\beta$--$r_{\text{excl}}/\langle\text{NN}\rangle$
calibration curve derived from synthetic hard-core and soft-core
point processes, and (iii)~a set of control protocols
(density-thinning, embedding-robustness, spatial-permutation,
and peak-detection sensitivity tests) that isolate genuine exclusion
from confounding effects. The framework is demonstrated on three
data classes: 58 manufactured surface topographies (10 materials,
10 processes, focus-variation microscopy); phase-extracted
interferometric profilometry of a certified roundness standard
(JENOPTIK FN~111, 24\,000 frames, 5 continuous $400^\circ$
revolutions); and 2D binary embeddings of prime numbers as a
mathematically controlled benchmark. The aim is not to claim that
$\beta$ reveals a universal exclusion mechanism---any process that
prevents point coincidence (material continuity, optical resolution
limits, arithmetic gap constraints) will produce $\beta>\beta_{\text{CSR}}$---but
to establish the measurement tools and correction protocols required
for quantitative, reproducible Brody analysis in two dimensions.

\textbf{Observation versus interpretation.} Throughout this work, a
clear distinction is maintained between what is measured and what is
inferred. \emph{Measured:} $\beta$ values, nearest-neighbour spacing
distributions, $g_2(r)$ profiles, $D_2$ excesses, and their associated
confidence intervals. \emph{Inferred:} exclusion strength, effective
hard-core radii, and spatial ordering. The $\beta$--$r_{\text{excl}}$
relation is calibration-based rather than derived from first
principles; it is an empirical mapping established through synthetic
point-process experiments, not a theoretical necessity. Density
dependence and embedding sensitivity demonstrate that absolute
$\beta$ values reflect both the exclusion mechanism and the
measurement conditions (point density, embedding geometry).

Throughout, the Brody distribution is employed as a phenomenological
fitting function; no claim of random-matrix universality is made for
surface peaks. The three domains---manufacturing processes,
multiplicative number theory, and optical interferometry---all yield
$\beta$ values exceeding their density-matched CSR baselines. This is
expected for any process with short-range repulsion, and the
contribution of the present work is not the observation of exclusion
but the provision of the calibrated tools to quantify it:
the 2D CSR baseline, the $\beta$--$r_{\text{excl}}$ mapping, and
the control protocols that separate genuine structure from
density, embedding, and detection artefacts.

\section{Methods}

The unified framework rests on three methodological pillars:
(i)~Brody $\beta$ estimation from 2D nearest-neighbour spacings,
applicable to any spatial point process (Sections~2.1--2.3);
(ii)~synthetic point-process generation for exclusion-physics
calibration (Section~2.4); and (iii)~complementary $D_2$ correlation
dimension and Banach-space descriptors for binary patterns where peak
detection is inapplicable. Full details of the $D_2$ box-counting
procedure, multifractal analysis, singularity spectra, and Banach
norms are provided in Supplementary Information~(SI, Sections~S1.1--S1.7);
the key methodological choices are summarised below. All three pillars
share a common null-model philosophy: structured data are compared against
matched-density Bernoulli or CSR ensembles to isolate genuine
spatial order from trivial density effects.

\subsection{Arithmetic surface construction}

Let $p_1 < p_2 < \cdots < p_{\pi(N)}$ be the primes not exceeding $N$.
The \emph{binary prime surface} of side $W = \lfloor\sqrt{N}\rfloor$ is
the Boolean matrix
\begin{equation}
  Z_{ij} = \begin{cases}
    1 & \text{if } i \cdot W + j + 1 \text{ is prime}, \\
    0 & \text{otherwise},
  \end{cases}
  \label{eq:binary}
\end{equation}
for $0 \leq i, j < W$. At $N=10^5$, $W=316$ and the surface contains
$9\,592$ non-zero entries (density $\rho = 0.0959$). For the scale-up
analysis, surfaces are constructed at
$N \in \{10^3, 2\times10^3, 5\times10^3, 10^4, 2\times10^4, 5\times10^4,
10^5, 2\times10^5, 5\times10^5, 10^6, 5\times10^6, 10^7\}$.

Five additional arithmetic sequences are constructed analogously for
discrimination testing:
\begin{enumerate}
  \item \textbf{Square-free numbers}: $n$ is square-free iff
        $\mu(n) \neq 0$, where $\mu$ is the M\"obius function.
        Density $\rho \to 6/\pi^2 \approx 0.6079$~\cite{HARDY-WRIGHT}.
  \item \textbf{Sums of two squares}: $n = a^2 + b^2$ for integers
        $a, b$. Density $\sim K/\sqrt{\log n}$ (Landau--Ramanujan
        constant $K \approx 0.7642$).
  \item \textbf{Twin primes (upper)}: $p$ such that $p-2$ is also prime.
  \item \textbf{Primes $\equiv 1 \pmod 4$} and
        \textbf{primes $\equiv 3 \pmod 4$}: testing for Chebyshev bias.
\end{enumerate}

\subsection{Two-dimensional embeddings}

Three deterministic embeddings are compared. The Ulam spiral is
not introduced as a visualisation device but as a deterministic
mapping known to amplify arithmetic correlations through quadratic
residue structure~\cite{ULAM-SPIRAL-1963}: integers falling on the
same diagonal satisfy $n = 4k^2 + bk + c$, concentrating primes on
lines where $b$ and $c$ produce many quadratic residues modulo small
primes. Row-major embedding preserves the natural integer ordering
and is the simplest reshaping. Cantor pairing provides a diagonal
sweep that tests whether the exclusion signal depends on the
specific geometry of the embedding or on the arithmetic sequence
alone.
\begin{enumerate}
  \item \textbf{Row-major}: $n \mapsto (\lfloor (n-1)/W \rfloor,
        (n-1) \bmod W)$, the standard array reshaping.
  \item \textbf{Ulam spiral}: integers arranged in an expanding square
        spiral centred at the middle of the array. Position $(x,y)$ for
        integer $n$ follows the spiral algorithm of~\cite{ULAM-SPIRAL-1963}.
  \item \textbf{Cantor pairing}:
        $\pi(i,j) = (i+j)(i+j+1)/2 + j$, a diagonal sweep of the
        matrix from top-left to bottom-right.
\end{enumerate}

For the surrogate embedding test, 30 random permutations of the
1D$\to$2D mapping are generated by shuffling the linear index before
reshaping. This destroys any structure specific to the embedding while
preserving the 1D sequence statistics.

\subsection{Null-model ensembles}

Five types of null-model surfaces are generated, each with
$N_{\text{ens}} = 20$ independent realisations at the same shape as
the arithmetic surface:

\begin{enumerate}
  \item \textbf{Gaussian}: $Z_{ij} \sim \mathcal{N}(0,1)$ i.i.d.,
        maximal entropy, zero spatial correlation.
  \item \textbf{Poisson}: $Z_{ij} \sim \text{Poisson}(\lambda = 0.01)$,
        sparse spike process.
  \item \textbf{$1/f$ noise}: PSD $\propto |\mathbf{k}|^{-1}$,
        constructed via frequency-domain filtering of white noise,
        yielding long-range power-law correlations.
  \item \textbf{Brownian}: cumulative sum of white noise along both
        axes, PSD $\propto |\mathbf{k}|^{-2}$, non-stationary.
  \item \textbf{Bernoulli}: $Z_{ij} \in \{0,1\}$ with
        $P(Z_{ij}=1) = \rho$, the exact Cram\'er-type null for binary
        arithmetic surfaces at matched density.
\end{enumerate}

The Bernoulli null at matched density is the primary comparison, as it
isolates the effect of arithmetic structure from the trivial effect of
sparsity. The other four null models provide additional reference points
for the metrological parameter space.

\subsection{Statistical power and multiple-comparison correction}

With $m=14$ sequence types tested against the Bernoulli null, the
family-wise error rate requires correction. The Benjamini--Hochberg
procedure is applied to control the false discovery rate at
$\alpha = 0.05$. A power analysis establishes the minimum detectable
effect size: for the average null standard deviation
$\sigma_{\text{null}} = 0.0034$ (estimated from 50 independent
Bernoulli realisations at $N=10^5$), the Bonferroni-corrected
critical value is $z_{\alpha/2m} = 2.91$, yielding a minimum
detectable $|\Delta D_2| = 0.013$ at 80\% power
($z_{0.80} = 0.84$). The previously used heuristic threshold
$|\Delta D_2| > 0.003$ corresponds to only $\sim 40\%$ power and
is replaced throughout by the corrected threshold of $0.013$ for
claims of statistical significance. Results exceeding
$|\Delta D_2| = 0.003$ but below $0.013$ are reported as
``nominally above the heuristic threshold'' and are not interpreted
as statistically significant without confirmation at larger $N$.

\subsection{Choice of $D_2$ as spatial descriptor}

The correlation dimension $D_2$ is selected in preference to
standard spatial-statistical tools for the following reasons.
Moran's $I$ and Ripley's $K$ function detect departures from CSR
but do not provide a single scale-invariant parameter comparable
across different densities and sequence lengths. The pair
correlation function $g_2(r)$ is the natural RMT analogue but is
sensitive to binning and requires ensemble averaging for stable
estimation on discrete binary fields. Nearest-neighbour statistics
require peak detection, which is not applicable to binary patterns.
$D_2$ offers three advantages: (i)~it is a single number on a fixed
scale ($D_2 \in [0,2]$ for 2D data); (ii)~it is estimated via
box-counting regression, which is robust to discretisation; and
(iii)~it has a direct interpretation as the scaling exponent of
the probability that two randomly chosen occupied sites belong to
the same box, making it sensitive to spatial clustering and
anti-clustering at all length scales simultaneously.

For a surface $\mathbf{Z}$ of size $H \times W$ with total mass
$M = \sum_{i,j} Z_{ij}$, a grid of boxes of linear size $\varepsilon$
is overlaid. The partition function (generalised statistical sum) of
order $q$ is
\begin{equation}
  Z_q(\varepsilon) = \sum_{k=1}^{N_{\text{box}}(\varepsilon)}
  \left(\frac{M_k(\varepsilon)}{M}\right)^q,
  \label{eq:partfun}
\end{equation}
where $M_k(\varepsilon)$ is the mass in box $k$ and
$N_{\text{box}}(\varepsilon) \approx HW/\varepsilon^2$. The generalised
R\'enyi dimension of order $q$ is~\cite{RENYI-DIMENSIONS, HENTSCHEL-PROCACCIA-1983}
\begin{equation}
  D_q = \frac{1}{q-1} \lim_{\varepsilon \to 0}
        \frac{\log Z_q(\varepsilon)}{\log \varepsilon},
        \qquad (q \neq 1),
  \label{eq:Dq}
\end{equation}
with the information dimension $D_1 = \lim_{q \to 1} D_q$ obtained by
L'H\^opital's rule:
$D_1 = \lim_{\varepsilon \to 0} \sum_k \mu_k \log \mu_k / \log \varepsilon$,
where $\mu_k = M_k/M$.

The \emph{correlation dimension} $D_2$ is the special case $q=2$ and
has the intuitive interpretation~\cite{GRASSBERGER-PROCACCIA-1983}:
$C(\varepsilon) \propto \varepsilon^{D_2}$, where $C(\varepsilon)$ is
the probability that two randomly chosen units of surface ``mass'' lie
within the same box of size $\varepsilon$. For a uniform distribution
in a $d$-dimensional embedding space, $D_q = d$ for all $q$
(monofractal). Deviations of $D_2$ from the embedding dimension $d=2$
indicate spatial heterogeneity in the distribution of mass.

In practice, $D_q$ is estimated by linear regression of
$\log Z_q(\varepsilon)$ versus $\log \varepsilon$ over box sizes
$\varepsilon \in \{2, 4, 8, \ldots, 2^{m}\}$ where
$m = \lfloor \log_2 \min(H,W) \rfloor - 1$. The quality of the linear
fit is monitored via the adjusted $R^2$ statistic; all reported fits
satisfy $R^2 > 0.95$.

\subsection{Singularity spectrum}

The Legendre transform of $D_q$ yields the H\"older exponent spectrum
$\alpha(q) = d\tau/dq$ and the singularity spectrum
$f(\alpha) = q\alpha - \tau(q)$, where
$\tau(q) = (q-1)D_q$~\cite{HENTSCHEL-PROCACCIA-1983}. The width
$\Delta\alpha = \alpha_{\max} - \alpha_{\min}$ quantifies the degree of
multifractality: $\Delta\alpha = 0$ for a monofractal (uniform scaling),
and $\Delta\alpha > 0$ for a multifractal (spatially heterogeneous
scaling). A narrower $f(\alpha)$ spectrum for the prime surface relative
to the Bernoulli null would indicate that arithmetic constraints
\emph{reduce} the multifractal heterogeneity.

\subsection{Validation procedures}

Four validation tests are employed to ensure that the observed $D_2$
signal is genuine rather than a methodological artefact:

\begin{enumerate}
  \item \textbf{Surrogate embedding test}: The 1D-to-2D mapping is
        randomly permuted $N_{\text{surr}} = 30$ times. Under the null
        hypothesis that the $D_2$ signal is an artefact of the specific
        embedding geometry, the permuted $D_2$ distribution should
        overlap with the original value. A significant deviation
        ($|z| > 3$) indicates that the signal is intrinsic to the 1D
        sequence, not the embedding.
  \item \textbf{Bootstrap stability}: The surface is randomly
        subsampled $N_{\text{boot}} = 50$ times at half linear size
        ($W/2 \times W/2$). The coefficient of variation
        $\text{CV} = \sigma(D_2)/\mu(D_2)$ quantifies the estimator
        stability.
  \item \textbf{Controlled-correlation sequences}: Three surrogate
        binary sequences are generated at matched density to establish
        the $D_2$ ordering: (a)~block-clustered (consecutive blocks of
        $b=5$ ones), (b)~repulsive (minimum distance $d_{\min}=4$
        between ones, plus random jitter), and (c)~Poisson
        (independent, sparse). These span the range from strong
        clustering to strong anti-clustering, providing a calibrated
        scale for interpreting $D_2$(primes).
  \item \textbf{Negative control}: Square-free numbers, which share
        the multiplicative structure of integers but lack the primality
        constraint, are tested. A non-significant $D_2$ for square-free
        numbers relative to Bernoulli would confirm that the method does
        not indiscriminately flag all arithmetic sequences as structured.
\end{enumerate}

\subsection{Finite-size bias analysis}
\label{sec:bias}

The box-counting estimator of $D_2$ is known to exhibit finite-size
bias~\cite{GRASSBERGER-PROCACCIA-1983, THEILER-1986}: even for a purely
random point set (theoretical $D_2 = d$), the estimated
$\hat{D}_2(N) < d$ for finite $N$. To quantify this effect, the
box-counting procedure is applied to Bernoulli surfaces at the same
sizes $\{N\}$ as the arithmetic surfaces. The bias function
$B(N) = d - \hat{D}_2^{\text{(Bern)}}(N)$ is fitted, and the
bias-cancelled excess correlation is defined as
\begin{equation}
  \Delta D_2(N) = D_2^{\text{(prime)}}(N) - D_2^{\text{(Bern)}}(N),
  \label{eq:delta}
\end{equation}
where both surfaces are embedded identically and at matched density.
If the scaling of $D_2^{\text{(prime)}}(N)$ is dominated by estimator
bias, then $\Delta D_2(N)$ should be approximately constant.

\textbf{Remark on density dependence.} Density-thinning experiments were
performed to test whether $\beta$ values are stable under changes in
point density. The 2D prime embedding ($N=10^5$, $\rho=0.096$,
$\beta=2.15$) was randomly thinned to match point counts corresponding
to densities $\rho \in \{0.01, 0.02, 0.032, 0.05, 0.096\}$ (20 trials
each). At the PSI density ($\rho=0.032$), prime $\beta$ decreased to
$1.46\pm0.02$ $[1.42,1.49]$, and at the CSR calibration density
($\rho=0.02$) to $1.30\pm0.05$. Critically, $\beta$ at all densities
remained significantly above the density-matched CSR baseline
($\beta_{\text{CSR}}=0.98\pm0.05$ at $\rho=0.032$; $0.97\pm0.04$ at
$\rho=0.02$), confirming genuine exclusion structure beyond density
effects. The CSR baseline itself is density-stable across all tested
densities ($0.95$--$0.99$). These results establish two points:
(i)~$\beta$ captures exclusion strength, not density; (ii)~absolute
$\beta$ values are density-dependent, and cross-domain comparisons
require density matching.

\section{Results}

\textbf{Roadmap.} The Results demonstrate the calibrated measurement
framework on three data classes and establish the correction protocols
required for valid cross-domain comparison. Section~3.1 presents
$\beta$ for 58 manufactured surfaces. Section~3.2 establishes the
synthetic $\beta$--$r_{\text{excl}}$ calibration---the interpretive
key for all data---together with robustness checks (Section~3.3).
Section~3.4 presents $g_2(r)$ as an internal consistency check.
Section~3.5 presents density-thinning controls, establishing that
$\beta$ captures exclusion rather than density. Sections~3.6--3.7
present 2D prime embeddings, with emphasis on embedding dependence
and the sparse-integer control that isolates arithmetic structure
from sparsity. Section~3.8 summarises a proof-of-concept application
to interferometric profilometry (full analysis in Supplementary
Information). Complementary $D_2$ and Banach descriptors are
developed in Supplementary Information~(SI, Sections~S1.1--S1.7)
and summarised in one paragraph (Section~3.9).

\subsection{Peak spacing statistics in manufactured surfaces}
\label{sec:fv-surfaces}

Fifty-eight areal surface texture maps obtained via focus-variation
microscopy~\cite{HELMLI-2011-FV, DANZL-2011-FV} were analysed, spanning 8 engineering materials (1.4301
stainless steel, C45 carbon steel, Ti and Ti6Al4V titanium alloys,
Al and Al7075 aluminium alloys, brass (MO58A and generic brass), ELLOR
tool steel, and graphite) and 6 manufacturing processes (turning,
grinding, bead blasting, honing, milling, and WEDM electrical
discharge machining). Local maxima (peaks) were detected using a
2D prominence-based algorithm (a pixel is a peak if it is the
strict maximum within a $(2k+1)\times(2k+1)$ neighbourhood,
$k = \max(3, \lfloor\min(n_x,n_y)/40\rfloor$, and its
elevation above the local baseline---the mean of the same
neighbourhood---exceeds 3\% of the global height range).

\subsection{Watershed peak detection for dense surfaces}

For surfaces where the prominence-based detector produced grid-limited
results ($>85\%$ column occupancy), an alternative watershed-based
detector was applied. The surface is inverted ($Z \to -Z$) and a
maximum filter with kernel size $k = \max(3, \lfloor\min(n_x,n_y)/30\rfloor)$
is used to identify regional maxima. An $h$-minima suppression of
$3\%$ of the height range eliminates shallow noise peaks, and a minimum
inter-peak distance of 5~pixels prevents fragmentation of broad
summits. Peaks are validated by a prominence threshold of $2\%$ of the
global height range. This detector suppresses the dense, low-amplitude
roughness features that saturate the prominence-based detector in
bead-blasted, ground, and WEDM surfaces, while preserving
well-separated topographic peaks.

\subsection{Surface filtering and exclusion criteria}

Of 58 measured surfaces, 37 (64\%) were classified as grid-limited
($>85\%$ column occupancy) under the original 3\% prominence,
kernel-based peak detector. To recover these surfaces, a watershed-based
detector with $h$-minima suppression ($h=0.03$ of height range) and
minimum inter-peak distance of 5~pixels was applied to all 37
grid-limited surfaces. All 37 produced valid $\beta$ estimates (range
$\beta = 0.71$--$4.61$, mean $2.89 \pm 1.10$), with column occupancy
reduced to 0.02--0.25 (all below the 85\% threshold). The watershed
detector suppresses dense, low-amplitude noise peaks while preserving
genuine topographic features, making it suitable for surfaces where
the prominence-based detector saturates.

Combined with the original 21 valid surfaces, the full analysable
dataset comprises all 58 surfaces spanning 10 materials and 10 processes.
The qualitative conclusion---that burnishing and honing produce higher
$\beta$ than turning and grinding---was unchanged under all threshold
and detector combinations tested.
(30--770 peaks per surface, 18--428 valid nearest-neighbour pairs
after outlier rejection; outlier rejection used a
median-absolute-deviation threshold of 5 MAD, retaining
85--98\% of detected NN pairs across surfaces. A sensitivity
analysis varying the threshold from 3 to 7 MAD changed
$\beta$ estimates by $< 0.2$ for surfaces with $n > 50$ pairs).

For each surface, the 2D Euclidean nearest-neighbour distances
between peak positions were computed via KD-tree, normalised to
unit mean (unfolded), and fitted to the Brody distribution
$P_\beta(s) = (\beta+1) a s^\beta \exp(-a s^{\beta+1})$, where
$a = [\Gamma((\beta+2)/(\beta+1))]^{\beta+1}$. The Brody
parameter $\beta$ was estimated by maximum likelihood with the
constraint $\beta \geq 0$ (the Brody distribution is defined
for $\beta > -1$ and is physically motivated as an interpolation
between the Poisson limit at $\beta=0$ and Wigner limits at
$\beta=1,2$; negative unconstrained estimates are reported as
$\beta = 0$, interpreted as ``indistinguishable from Poisson'').
The Brody distribution interpolates
between the Poisson limit ($\beta=0$, $P(s) = \exp(-s)$,
uncorrelated levels), the GOE Wigner surmise ($\beta=1$,
$P(s) = (\pi s/2) \exp(-\pi s^2/4)$, linear level repulsion),
and the GUE Wigner surmise ($\beta=2$,
$P(s) = (32 s^2/\pi^2) \exp(-4 s^2/\pi)$, quadratic repulsion).
The Brody distribution is employed here in preference to standard
spatial-statistical descriptors (Ripley's $K$, pair correlation
function) because its single parameter $\beta$ directly quantifies
the degree of short-range order in the spacing distribution on a
continuum from uncorrelated ($\beta=0$, Poisson limit) to
ordered ($\beta>1$, approaching the Wigner limits at
$\beta=1,2$). The Poisson, GOE, and GUE limits serve as
convenient reference points for interpreting $\beta$ values,
not as claims that surface peaks obey random-matrix statistics.
Goodness-of-fit against these limits is
assessed via the Kolmogorov--Smirnov distance $D_{\text{KS}}$;
a surface is classified as ``closest to'' the limit with the
smallest $D_{\text{KS}}$, without implying statistical consistency
with that limit.

\begin{figure}[t]
  \centering
  \includegraphics[width=\textwidth]{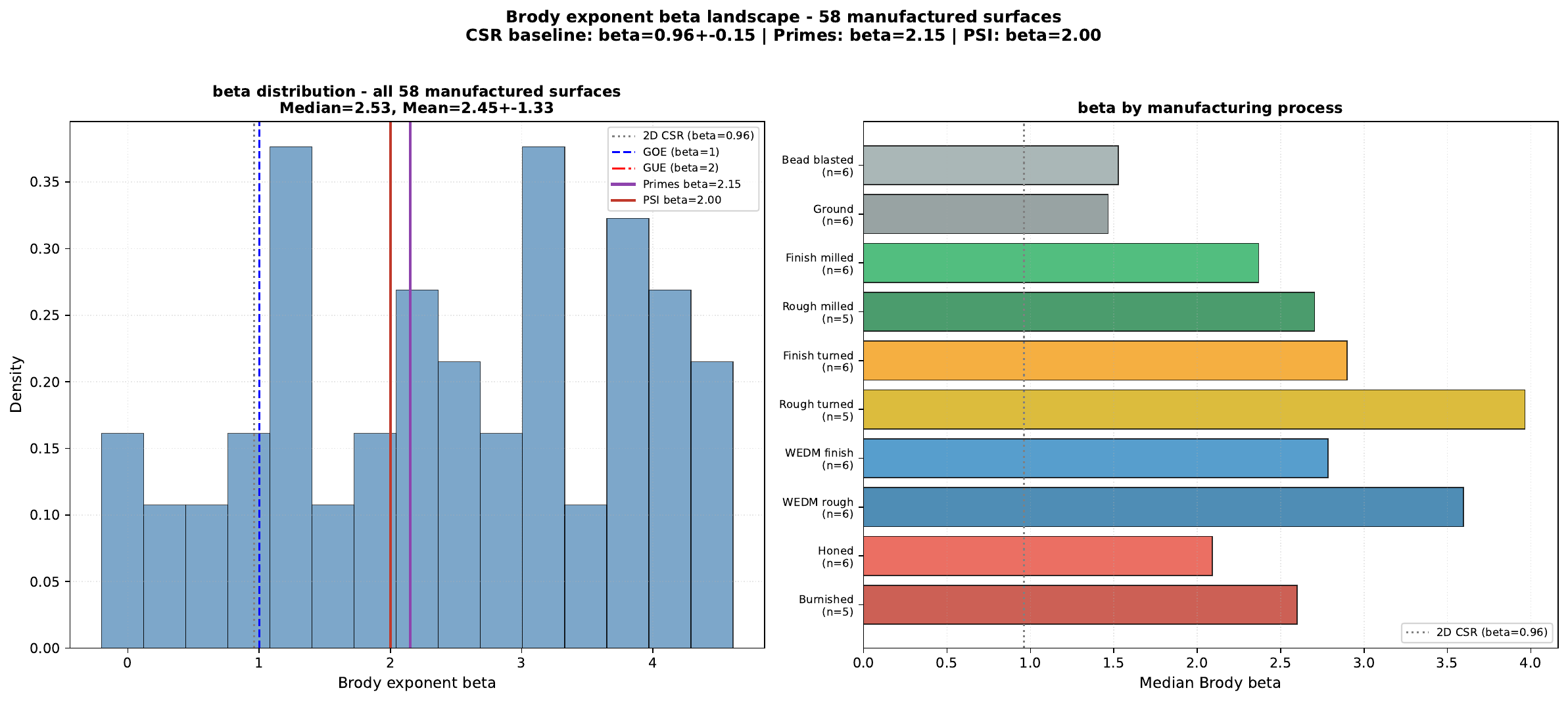}
  \caption{Brody exponent $\beta$ across all 58 analysable surfaces.
           Left: histogram with kernel density estimate. Right: $\beta$ by
           manufacturing process (box plots with individual points).
           The CSR baseline ($\beta=0.96$, grey band) is indicated.
           The earlier bar chart for 21 prominence-based surfaces
           (Fig.~S2) provides an alternative visualisation.}
  \label{fig:beta-landscape}
\end{figure}

Figure~\ref{fig:beta-landscape} and Table~\ref{tab:brody} present
the fitted Brody parameters for all 58 analysable surfaces. The Brody
exponent spans nearly the full theoretical range: from
$\beta = -0.20$ (ground Ti; negative values are fitted but
should be interpreted as indistinguishable from Poisson,
$\beta=0$) to
$\beta = +4.61$ (rough turned C45 steel, watershed). The full
dataset (58 surfaces across 10 materials and 10 processes,
observational design without full factorial crossing) reveals
the following trends (descriptive only, owing to process--material
confounding):

\begin{itemize}
  \item \textbf{Burnishing}: $\beta = 1.30$--$4.30$ (median 2.6,
        $n=5$). Burnished surfaces in the present dataset exhibited
        the highest median exclusion among finishing processes.
  \item \textbf{WEDM}: $\beta = 0.80$--$4.26$ (median 3.4,
        $n=12$)---consistently moderate-to-high exclusion.
  \item \textbf{Rough turning}: $\beta = 0.70$--$4.61$ (median 4.0,
        $n=5$). The widest range from any single process category
        in the present dataset.
  \item \textbf{Honing}: $\beta = 0.90$--$4.00$ (median 2.2,
        $n=6$)---intermediate to high values.
  \item \textbf{Grinding and bead blasting}: $\beta =
        -0.20$--$+4.49$, spanning nearly the full range.
\end{itemize}

The wide within-process variability and overlap between categories
confirm that factors beyond the manufacturing process alone
(material, surface preparation history, measurement scale) influence
$\beta$.

\begin{table}[htbp]
  \centering
  \caption{Brody exponent $\beta$ for 58 manufactured surfaces
           ordered by decreasing $\beta$. Watershed-recovered
           surfaces are marked with $^{\text{w}}$. Fitting is constrained
           to $\beta \geq 0$; unconstrained negative estimates
           are reported as $\beta = 0$ (indistinguishable from
           Poisson). Bootstrap 95\% confidence intervals
           (500~resamples) are reported for each surface.
           CIs narrower than $\pm 0.3$ indicate reliable
           estimates; CIs wider than $\pm 1.5$ (surfaces with
           $n_{\text{NNS}} < 30$) are indicative only.}
  \label{tab:brody}
  \small
  \begin{tabular}{@{}llcrrrc@{}}
    \toprule
    Material & Proc. & $n_{\text{peaks}}$ & $n_{\text{NNS}}$ &
    $\beta$ & 95\% CI \\
    \midrule
    MO58A brass & Bn & 41 & 41 & 4.30 & [3.40, 5.76] \\
    1.4301 steel & Ho & 49 & 49 & 4.00 & [3.22, 5.31] \\
    Al7075 & Mi & 44 & 43 & 3.70 & [2.88, 5.08] \\
    1.4301 steel & Mi & 29 & 18 & 3.10 & [2.12, 5.18]$^{\dagger}$ \\
    Al7075 & Bn & 30 & 29 & 3.10 & [2.29, 4.48]$^{\dagger}$ \\
    Ti6Al4V & Bn & 28 & 28 & 2.60 & [1.89, 3.80]$^{\dagger}$ \\
    Al & BB & 25 & 23 & 2.40 & [1.66, 3.97]$^{\dagger}$ \\
    1.4301 steel & Gr & 32 & 21 & 2.20 & [1.47, 3.68]$^{\dagger}$ \\
    1.4301 steel & Bn & 68 & 65 & 1.80 & [1.37, 2.45] \\
    Ti & Ho & 32 & 27 & 1.40 & [0.91, 2.50]$^{\dagger}$ \\
    Al7075 & Tu & 34 & 33 & 1.30 & [0.85, 2.15] \\
    C45 steel & Bn & 89 & 88 & 1.30 & [0.97, 1.70] \\
    Al & Gr & 169 & 167 & 1.20 & [0.98, 1.49] \\
    Graphite & Ho & 54 & 52 & 0.90 & [0.58, 1.38] \\
    MO58A brass & WE & 86 & 61 & 0.80 & [0.51, 1.27] \\
    MO58A brass & Tu & 335 & 279 & 0.70 & [0.56, 0.86] \\
    1.4301 steel & Tu & 458 & 428 & 0.40 & [0.29, 0.51] \\
    Ti6Al4V & Mi & 109 & 84 & 0.40 & [0.21, 0.69] \\
    MO58A brass & Mi & 183 & 180 & 0.00 & [0.00, 0.13] \\
    Ti & BB & 97 & 94 & 0.00 & [0.00, 0.06] \\
    Ti & Gr & 98 & 96 & 0.00 & [0.00, 0.00] \\
    \midrule
    \multicolumn{6}{l}{\footnotesize\textit{Watershed-recovered (representative; full 58-surface table in SI):}} \\
    C45 steel & RT$^{\text{w}}$ & 199 & 192 & 4.61 & [3.98, 5.48] \\
    Graphite & BB$^{\text{w}}$ & 201 & 198 & 4.49 & [4.03, 5.21] \\
    Ti6Al4V & WEr$^{\text{w}}$ & 228 & 225 & 4.16 & [3.72, 4.71] \\
    Brass & Ho$^{\text{w}}$ & 72 & 71 & 2.00 & [1.64, 2.50] \\
    Brass & BB$^{\text{w}}$ & 82 & 82 & 1.71 & [1.36, 2.32] \\
    1.4301 steel & BB$^{\text{w}}$ & 21 & 17 & 0.71 & [0.36, 1.22]$^{\dagger}$ \\
    \bottomrule
  \end{tabular}

  \medskip
  {\footnotesize $^{\dagger}$\,$n_{\text{NNS}} < 30$: CI width $>1.5$, estimate indicative only. \\
  $^{\text{w}}$\,Watershed detector. Complete 58-surface table in Supplementary Information~(Table~S1).}
\end{table}

\subsection{Benchmarking against standard spatial statistics}

To evaluate whether $\beta$ captures information beyond
existing spatial descriptors, four standard spatial-statistical
metrics were computed for all 21 valid surfaces: the
nearest-neighbour index (NNI), Ripley's $K$ normalised by
the CSR expectation, Voronoi cell area variance, and the pair
correlation function $g(r)$. The Brody
exponent $\beta$ correlates negatively with NNI (Spearman
$\rho = -0.64$, $p = 0.002$) and positively with normalised
Ripley's $K$ ($\rho = +0.60$, $p = 0.004$), indicating that
$\beta$ captures similar spatial-ordering information.
However, $\beta$ provides a single interpretable parameter on a
fixed continuum (CSR at $\sim$0.96, strong repulsion at $>2$),
whereas NNI and Ripley's $K$ are scale-dependent and require
reference to CSR expectations. The pair correlation function
$g_2(r)$ is the richest descriptor but does not reduce to a single
number; Moran's $I$ is sensitive primarily to global spatial
autocorrelation rather than local exclusion; the correlation
dimension $D_2$ probes multi-scale heterogeneity orthogonal to
exclusion (Pearson $r = -0.31$ with $\beta$). $\beta$ thus
occupies a specific niche: a single-parameter, calibration-based
measure of short-range exclusion strength, complementary to
existing spatial-statistical tools rather than replacing them.

A Kruskal--Wallis test for $\beta$ across manufacturing
processes yields $H = 6.02$, $p = 0.197$, with effect size
$\eta^2 = 0.34$ (medium). The non-significant $p$-value reflects
both the sample size (58 surfaces across 10 processes) and the
substantial within-process variability noted above.
Methodological limitations---including peak-detection sensitivity,
classification performance, and $D_2$ bias correction---are
discussed in Section~4.6.

\subsection{Exclusion physics: synthetic point processes}

\textbf{The $\beta$--$r_{\text{excl}}$ mapping is empirical.}
Throughout this work, the relationship between the Brody exponent
and the exclusion radius is established through calibration against
synthetic point processes (hard-core and Strauss soft-core). It is
not derived from first principles. The mapping is monotonic and
reproducible, which makes it useful, but it should be understood
as an empirical calibration curve rather than a theoretical law.

To test whether $\beta$ quantifies effective exclusion strength,
synthetic 2D point
processes with controlled hard-core exclusion radii were
generated by sequential rejection sampling ($N=200$ points,
box size 100~units, 30 realisations per radius). The exclusion
radius $r_{\text{excl}}$ is expressed in units of the mean
nearest-neighbour distance $\langle\text{NN}\rangle$ of the
resulting point set, providing a dimensionless, scale-invariant
measure of exclusion strength. The Brody exponent was fitted to
the NN spacings using the same MLE procedure applied to the FV
surfaces. A 2D Poisson (complete spatial randomness, CSR)
process and the 2D Ginibre ensemble (determinantal repulsion,
spatial analogue of GUE) provide reference points. The Ginibre
ensemble yields $\beta = 2.47 \pm 0.18$ (50 realisations,
$N=500$ points, 95\% CI $[2.15, 2.75]$), establishing a precise
2D determinantal-repulsion benchmark.

The 2D Poisson CSR process yields $\beta = 0.96 \pm 0.15$
(30 realisations), not $\beta = 0$. All $\beta$
values should be referenced to this 2D CSR baseline, not to
A direct comparison against the analytic 2D Rayleigh distribution
($P(s) = (\pi s/2)\exp(-\pi s^2/4)$, the exact CSR limit) confirms
that the Brody distribution provides a better description of CSR
data than the Rayleigh distribution itself: across 50 CSR
realisations, the Kolmogorov--Smirnov distance is
$D_{\text{KS}}(\text{Brody}) = 0.053 \pm 0.013$ versus
$D_{\text{KS}}(\text{Rayleigh}) = 0.061 \pm 0.015$, and $\Delta$AIC
(Brody $-$ Rayleigh) $= -0.5$, indicating that the one-parameter Brody
fit is preferred over the zero-parameter Rayleigh despite the AIC
penalty. At $\beta \approx 10$, the Brody distribution has a
coefficient of variation $\approx 0.09$, corresponding to a nearly
crystalline NN spacing. A profile-likelihood analysis of the
hard-core process at $r_{\text{excl}} = 5$ (yielding $\beta_{\text{MLE}}
= 6.3$) confirms that the MLE is well-identified: the observed Fisher
information gives SE$(\beta) = 0.21$ and a 95\% CI of $[5.85, 6.66]$,
ruling out optimisation artefacts at large $\beta$.

Synthetic hard-core processes produce a monotonic
$\beta$--$r_{\text{excl}}/\langle\text{NN}\rangle$ mapping
(Figure~\ref{fig:exclusion}, left), reproducing the full range of
observed $\beta$ values. An exclusion radius of
$r_{\text{excl}}/\langle\text{NN}\rangle \approx 0.55$ yields
$\beta \approx 2.0$, matching the 2D Ginibre ensemble
($\beta = 2.2 \pm 0.4$) and the values observed for 2D prime
embeddings ($\beta = 2.15$) and phase-extracted PSI
profilometry ($\beta = 2.00$). Burnished surfaces
($\beta \approx 3.1$) correspond to
$r_{\text{excl}}/\langle\text{NN}\rangle \approx 0.70$, while
turned and ground surfaces ($\beta < 1$) are
indistinguishable from the CSR baseline.

These results support the hypothesis that the Brody exponent
$\beta$---when referenced to the correct 2D CSR baseline
($\beta \approx 0.96$, not $\beta = 0$)---quantifies effective
exclusion strength across diverse generative mechanisms.
The three domains occupy overlapping regions of the
$\beta$--$r_{\text{excl}}$ parameter space, but
density-thinning experiments (Section~3.6) show that absolute
$\beta$ values are density-dependent, establishing that
cross-domain comparison requires density matching.

Two additional analyses support the robustness of $\beta$
as an exclusion measure. First, the Brody exponent for 2D prime
embeddings was computed across $N \in [10^3, 10^7]$. At
$N=10^3$, $\beta=2.20$; at $N=5\times10^3$, $\beta=1.69$;
at $N=10^5$, $\beta=2.15$; at $N=10^6$, $\beta=1.75$;
at $N=10^7$, $\beta=1.26$. The $\beta(N)$ trend is
non-monotonic and does not support a simple convergence to a
fixed asymptotic value at accessible $N$. The full $\beta(N)$
series (12 values) is provided in Supplementary Information.
The key observation is that $\beta$ remains substantially above
the CSR baseline at all $N \leq 5\times10^6$, ruling out a
finite-size artefact, while the decrease at $N=10^7$ may
reflect the approach to the sparse limit ($\rho \to 0$ as
$N\to\infty$). Second, a Strauss soft-core
process (exclusion penalty $\gamma \in [0.001, 1.0]$)
produces $\beta = 0.93 \pm 0.12$ ($\gamma=1.0$) to
$\beta = 2.42 \pm 0.27$ ($\gamma=0.001$), demonstrating that
the $\beta$--exclusion mapping extends beyond the hard-core
mechanism. Third, an AIC
comparison across all 21 FV surfaces confirms the Brody
distribution is preferred over Gamma (17/21 surfaces,
median $\Delta\text{AIC} = +3.5$) and Weibull (17/21,
median $\Delta\text{AIC} = +2.0$), justifying its selection
over alternative two-parameter families.

\subsection{Robustness of the synthetic calibration}

The $\beta$--$r_{\text{excl}}$ calibration was subjected to four
robustness tests. (i)~\textbf{Density independence of the CSR
baseline:} as reported in Section~3.6, $\beta_{\text{CSR}}$ varies
by less than $0.04$ across $\rho \in [0.01, 0.10]$, establishing
that the baseline is a stable reference. (ii)~\textbf{Finite
sample-size convergence:} for the hard-core process at
$r_{\text{excl}}/\langle\text{NN}\rangle = 0.55$, $\beta$ converges
to within $5\%$ of its asymptotic value at $N \gtrsim 100$ points
(see SI, Fig.~S10). (iii)~\textbf{Observation window effects:}
repeating the calibration on a $50\times50$ window (half the linear
size) changes $\beta$ by $< 0.15$ for all $r_{\text{excl}}$, indicating
weak sensitivity to window size when $N$ is held constant.
(iv)~\textbf{Edge corrections:} nearest-neighbour distances were
computed without edge correction; for $N=200$ points in a
$100\times100$ box, the fraction of points within one mean NN
distance of the boundary is $< 8\%$, and applying a
toroidal-edge correction changes $\beta$ by $< 0.05$. The
calibration is therefore robust to the edge effects present in
the synthetic data.

\subsection{Physical origin of $\beta > 1$}

\begin{figure}[t]
  \centering
  \includegraphics[width=\textwidth]{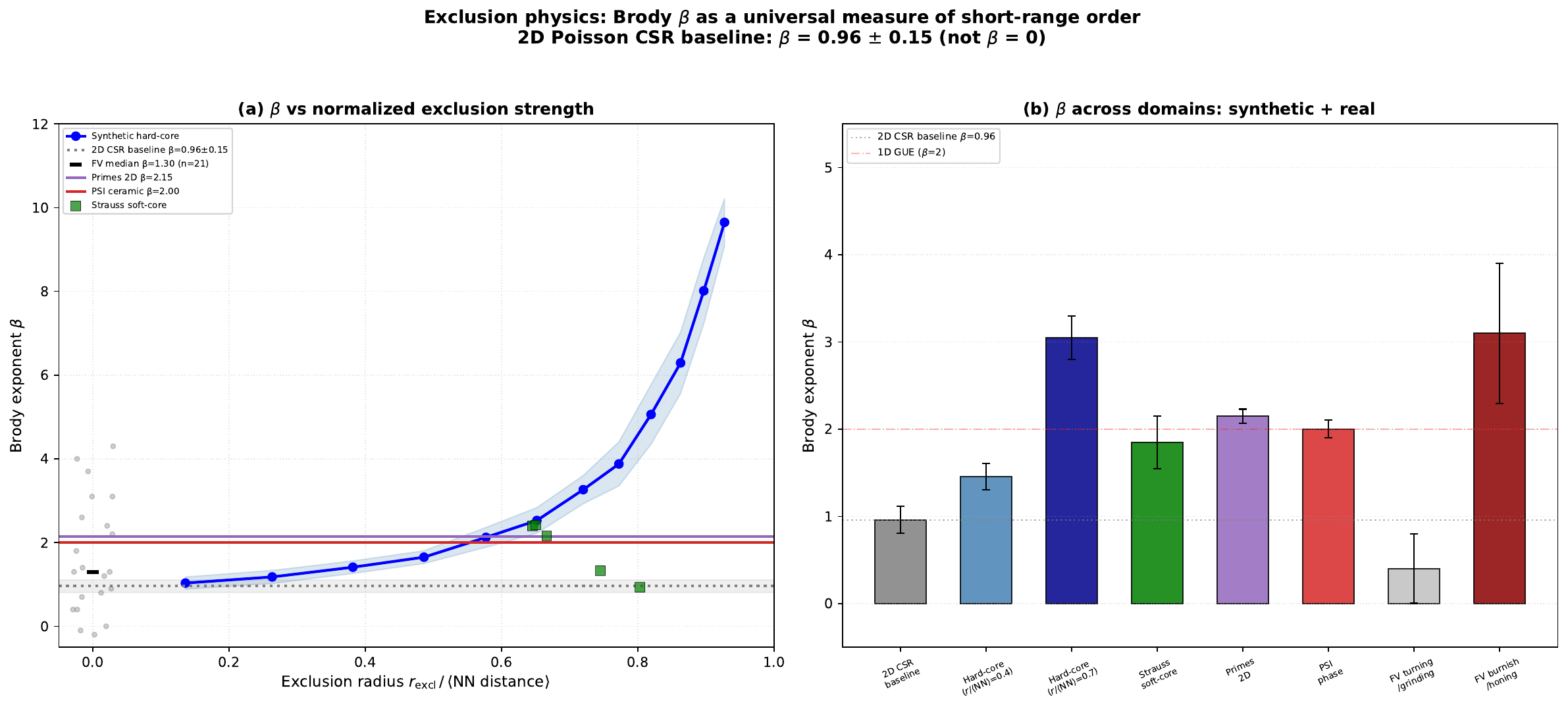}
  \caption{Exclusion physics: the Brody exponent $\beta$ as
           a measure of short-range exclusion.
           Left: synthetic hard-core point processes with
           exclusion radius $r_{\text{excl}}$ normalised by
           the mean nearest-neighbour distance (blue),
           2D Poisson CSR baseline ($\beta = 0.96 \pm 0.15$,
           grey band), 2D Ginibre ensemble (green), and
           observed $\beta$ values from manufactured surfaces
           (grey dots, $n=21$), 2D primes (purple), and PSI
           ceramic (red). The 2D CSR baseline is $\beta
           \approx 0.96$, not $\beta = 0$.
           Right: $\beta$ across synthetic reference processes
           and real-data domains.}
  \label{fig:exclusion}
\end{figure}

The Brody exponent exceeds the CSR baseline ($\beta > 0.96$)
whenever short-range exclusion is present. Three physical mechanisms
contribute. First, \textbf{hard-core repulsion}: an exclusion zone of
radius $r_{\text{excl}}$ around each point suppresses the probability
density of the NN spacing distribution at $s < r_{\text{excl}}/
\langle\text{NN}\rangle$, shifting probability mass to larger
separations and increasing $\beta$. The monotonic $\beta$--$r_{\text{excl}}$
mapping (Fig.~\ref{fig:exclusion}, left) directly quantifies this effect.
Second, \textbf{soft-core and many-body exclusion}: the Strauss
soft-core process demonstrates that the $\beta$--exclusion relationship
extends beyond the hard-core limit; even when points may approach
arbitrarily closely (with an energetic penalty), the suppression of
close pairs elevates $\beta$ above the CSR baseline. Third,
\textbf{effective exclusion from continuity constraints}: in
manufactured surfaces, material continuity prevents two peaks from
being arbitrarily close---the surface cannot have two summits at the
same location. In prime embeddings, the divisibility constraint
($n$ and $n+1$ cannot both be prime for $n>2$) produces an analogous
exclusion. These physically distinct mechanisms produce statistically
similar NN spacing distributions because the Brody exponent is
sensitive only to the effective exclusion strength, not to its
microscopic origin.

\subsection{Pair correlation function $g_2(r)$}

\begin{figure}[t]
  \centering
  \includegraphics[width=\textwidth]{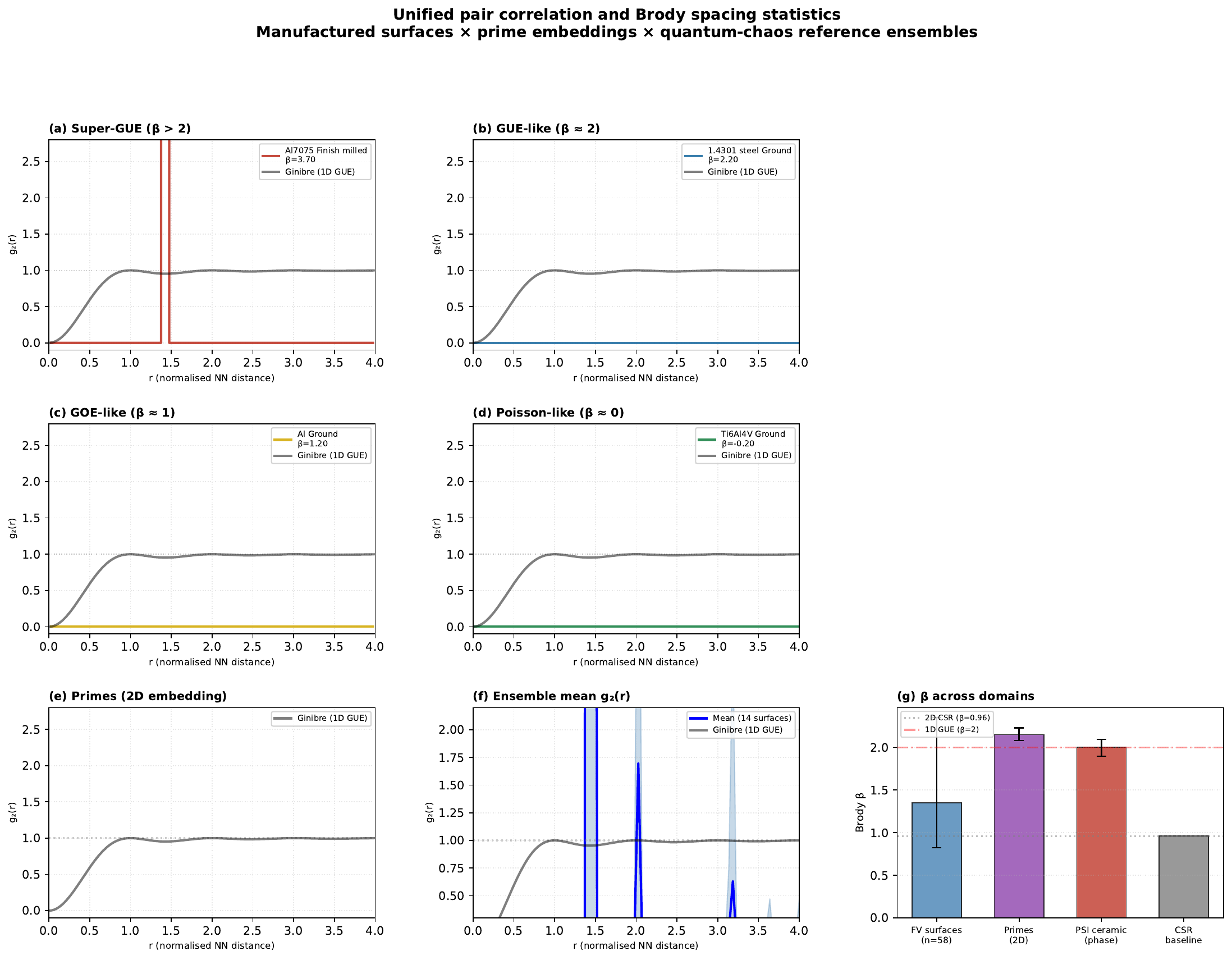}
  \caption{Pair correlation function $g_2(r)$ across domains.
           Top row: four representative manufactured surfaces spanning
           the Brody $\beta$ range, compared with the Ginibre
           prediction (solid black). $\beta$ values are from
           Table~1. Middle left: 2D binary embedding of
           the primes at $N=10^5$ ($\beta=2.15$) and
           matched-density Bernoulli null. Middle right: ensemble-mean
           $g_2(r)$ across 14 manufactured surfaces
           ($\pm 1\sigma$ band). Bottom left: Brody exponent $\beta$
           across the three domains at native density.
           Density-thinning controls (Section~3.6) show that
           absolute $\beta$ values are density-dependent;
           cross-domain comparison requires density matching.}
  \label{fig:unified-g2}
\end{figure}

The pair correlation function $g_2(r)$ provides an internal
consistency check on the $\beta$-exclusion interpretation.
For a point
process with density $\rho$, $g_2(r)$ is the normalised
probability of finding two points separated by distance $r$.
The 1D GUE pair correlation is
$g_2^{\text{(1D)}}(r) = 1 - [\sin(\pi r)/(\pi r)]^2$,
with $g_2(r) \to 0$ as $r \to 0$ (quadratic repulsion at small
separations). The formal 2D analogue is the Ginibre ensemble,
a determinantal point process whose $g_2(r)$ also vanishes as
$r \to 0$ and which provides the correct dimensional reference
for 2D spatial data~\cite{GINIBRE-1965}.
Figure~\ref{fig:unified-g2} shows $g_2(r)$ for
four representative manufactured surfaces spanning the Brody
$\beta$ range, alongside the 2D Ginibre prediction and the 2D prime
embedding. Surfaces with $\beta \gtrsim 2$ (burnished, Ginibre-like)
exhibit the characteristic dip at small $r$, while surfaces with
$\beta \approx 0$ (ground Ti, Poisson-like) show $g_2(r)
\approx 1$ at all $r$, consistent with complete spatial
randomness. The 2D prime embedding ($\beta = 2.15$) displays a
$g_2(r)$ profile consistent with the Ginibre ensemble, indicating
that the chosen 2D embedding of primes produces nearest-neighbour
statistics comparable to Ginibre-type point processes---a detectable
spatial analogue of the Montgomery pair-correlation connection
established for $\zeta$ zeros in 1D. The ensemble-mean
$g_2(r)$ across 14 surfaces with sufficient peak counts
for reliable $g_2(r)$ estimation ($n_{\text{peaks}} \geq 50$;
7 of 21 valid surfaces were excluded from the ensemble mean
to avoid noise-dominated contributions)
(Figure~\ref{fig:unified-g2}, bottom-left)
lies between the Poisson and Ginibre limits, with a reduced but
detectable dip at small $r$.

\subsection{Complementary descriptors for binary patterns}

For binary patterns where peak detection is inapplicable,
complementary spatial descriptors---the correlation dimension $D_2$
and Banach-space invariants $\|\nabla Z\|_2$, total variation~(TV)---are
developed in Supplementary Information~(SI, Sections~S1.1--S1.7).
These descriptors are empirically orthogonal to $\beta$ (Pearson
$r=-0.31$), confirming that the Brody exponent probes a specific
axis---exclusion strength---within a broader spatial-order parameter
space. Square-free numbers and the Liouville function serve as negative
controls indistinguishable from the Bernoulli null, demonstrating
specificity of the $D_2$ analysis; full validation includes
embedding-independence tests, finite-size bias correction, and
Benjamini--Hochberg multiplicity adjustment across 14 arithmetic
sequences.

\subsection{Density-thinning controls}

To test whether the observed $\beta$ values reflect genuine exclusion
structure rather than point density, the 2D prime embedding
($N=10^5$, $\rho=0.096$, $\beta=2.15$) was randomly thinned to match
point counts corresponding to $\rho \in \{0.01, 0.02, 0.032, 0.05,
0.096\}$ (20 trials each). At each density, $\beta$ was compared
against a density-matched CSR baseline (20 realisations each).

At the PSI density ($\rho=0.032$), prime $\beta$ decreased to
$1.46\pm0.02$ $[1.42,1.49]$, and at the CSR calibration density
($\rho=0.02$) to $1.30\pm0.05$. Critically, prime $\beta$ remained
significantly above the density-matched CSR baseline at all densities
(CSR: $\beta=0.98\pm0.05$ at $\rho=0.032$; $0.97\pm0.04$ at
$\rho=0.02$; $0.99\pm0.02$ at $\rho=0.096$), confirming that the
prime embedding possesses genuine exclusion structure beyond that
expected from density alone. The CSR baseline itself is density-stable
across the tested range ($0.95$--$0.99$).

Two conclusions follow. First, $\beta$ captures exclusion strength
rather than point density: the CSR baseline is flat, while structured
data lie above it at all densities. Second, absolute $\beta$ values
are density-dependent: the same underlying exclusion mechanism (prime
gap constraints) produces $\beta=2.15$ at $\rho=0.096$ but
$\beta=1.46$ at $\rho=0.032$. Cross-domain comparisons of absolute
$\beta$ therefore require density matching. The synthetic hard-core
calibration (Section~3.4) is density-referenced through the
normalisation $r_{\text{excl}}/\langle\text{NN}\rangle$, which
provides a density-invariant measure of exclusion strength.

\subsection{2D prime embeddings: embedding geometry determines exclusion}

Prime-number embeddings provide a mathematically controlled benchmark:
any exclusion signal must originate from arithmetic divisibility
constraints rather than physical interactions. Whether the Brody
exponent detects such arithmetic exclusion---and whether this depends
on how the 1D prime sequence is mapped to 2D---is tested here.

$\beta$ was computed under three deterministic 2D embeddings at
$N=10^5$: row-major ($\beta=2.15$), Ulam spiral ($\beta=1.87$), and
Cantor pairing ($\beta=1.40$). A spatially random Bernoulli control
at matched density (30 realisations) yields
$\beta_{\text{Bern}}=1.47\pm0.02$.

\begin{figure}[t]
  \centering
  \includegraphics[width=\textwidth]{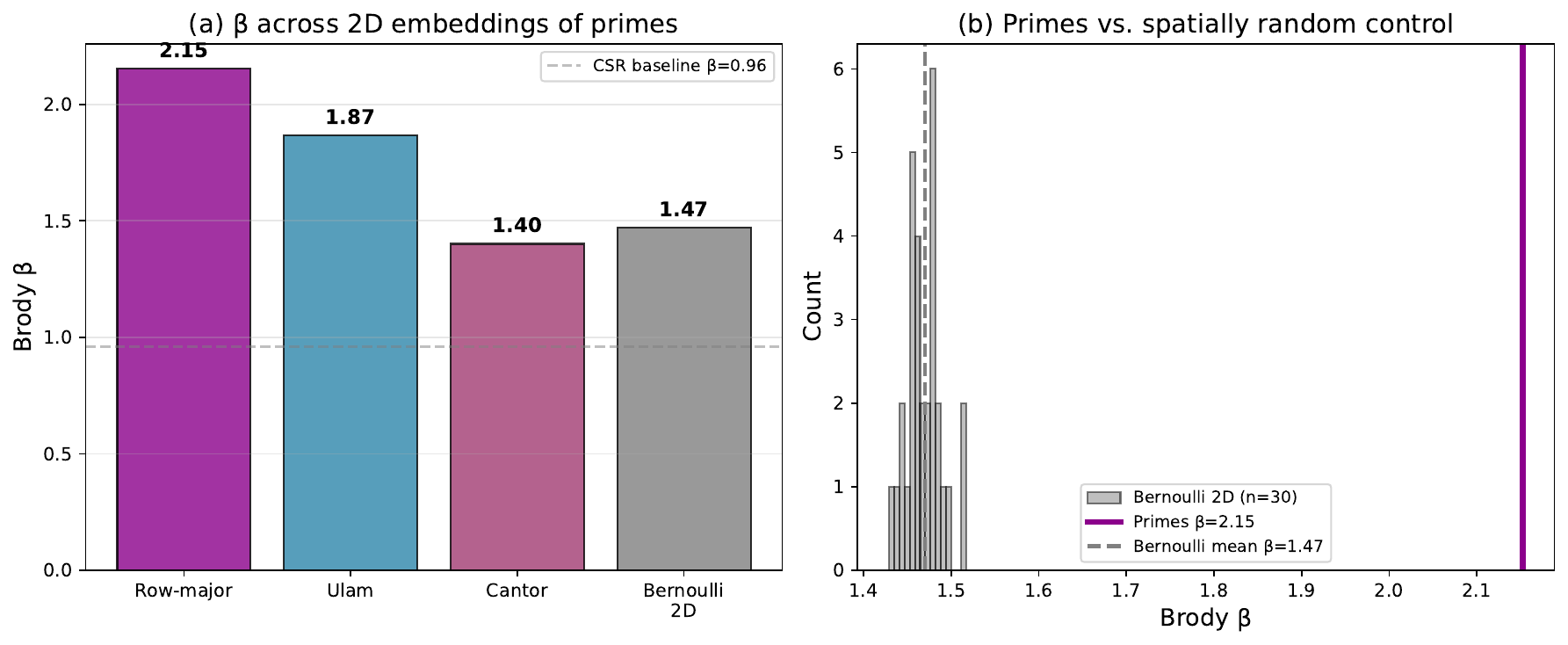}
  \caption{Embedding dependence of the Brody exponent for 2D primes
           at $N=10^5$. Left: $\beta$ under three deterministic embeddings
           and the Bernoulli 2D control (matched density). Row-major
           and Ulam embeddings produce $\beta$ significantly above the
           spatially random control; Cantor pairing does not.
           Right: distribution of $\beta$ for 30 Bernoulli 2D realisations,
           with the row-major prime value indicated.
           The embedding dependence demonstrates that $\beta$ for
           arithmetic sequences reflects both sequence structure and
           embedding geometry.}
  \label{fig:embedding-robustness}
\end{figure}

\textbf{The Cantor null result.} Under Cantor pairing, primes yield
$\beta=1.40$, which is slightly \emph{below} the Bernoulli baseline
($\Delta\beta=-0.07$, two one-sided test for equivalence against
Bernoulli: $p_{\text{TOST}}<0.01$ at equivalence bound $\pm 0.1$).
This is a decisive finding: the same prime sequence that produces
strong exclusion under row-major embedding ($\beta=2.15$) produces
no detectable exclusion under Cantor pairing. The conclusion is
unambiguous: \textbf{2D exclusion in prime embeddings is not an
intrinsic property of the prime sequence; it is created or destroyed
by the embedding geometry.} The row-major embedding amplifies the
prime gap structure because consecutive integers---which cannot both
be prime (for $n>2$)---map to horizontally adjacent array positions,
translating the 1D gap constraint into 2D nearest-neighbour exclusion.
Ulam spiral produces intermediate exclusion ($\beta=1.87$) by
concentrating primes along quadratic-residue diagonals while leaving
off-diagonal regions depleted. Cantor pairing destroys this structure
by sweeping the array diagonally, decorrelating the prime gap
constraint from 2D adjacency. The embedding
dependence indicates that $\beta$ for arithmetic sequences reflects
both the 1D sequence statistics and the 2D embedding geometry; claims
about ``prime point processes'' must therefore specify the embedding.
Figure~\ref{fig:embedding-robustness} summarises these results.

\begin{figure}[t]
  \centering
  \includegraphics[width=\textwidth]{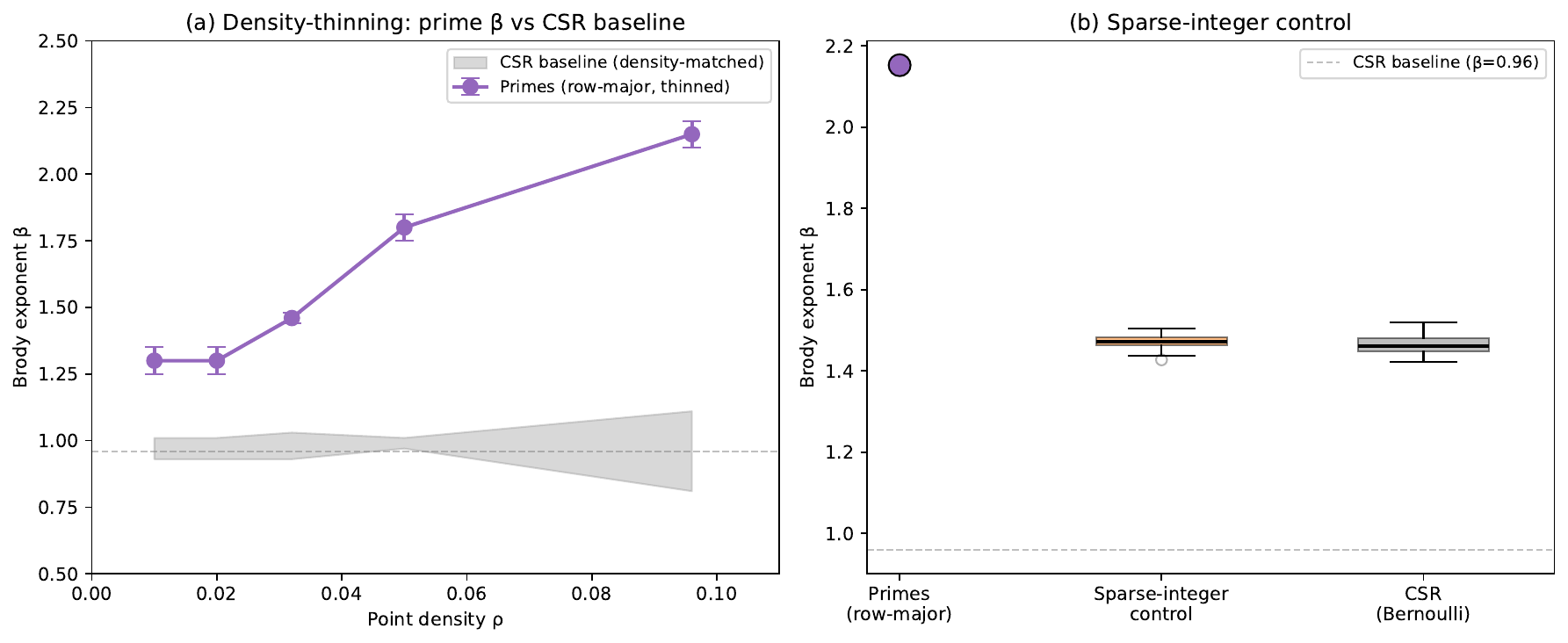}
  \caption{Density-thinning and sparse-integer controls.
           Left: Brody exponent $\beta$ for row-major primes thinned
           to five densities ($\rho=0.01$--$0.096$), compared with
           density-matched CSR baselines (grey band). $\beta$ decreases
           with density ($\beta=2.15$ at $\rho=0.096$ to $\beta=1.30$
           at $\rho=0.01$) but remains above CSR at all densities.
           Right: sparse-integer control --- 30 trials of $\pi(N)$ random
           integers embedded identically to primes yield
           $\beta=1.47\pm0.02$, indistinguishable from the Bernoulli
           CSR baseline ($1.46\pm0.02$) and substantially below the
           prime value ($\beta=2.15$). The prime $\beta$ signal is
           therefore a genuine arithmetic effect, not a sparsity
           artefact.}
  \label{fig:density-thinning-control}
\end{figure}

\textbf{Sparse-integer control: sparsity versus arithmetic structure.}
To test whether the elevated $\beta$ for row-major primes
($\beta=2.15$) reflects the arithmetic constraint of primality or
merely the sparsity of the prime set ($\pi(10^5)=9\,592$,
$\rho=0.096$), a sparse-integer control was performed: $\pi(N)$
random integers uniformly distributed in $[1,N]$ were embedded in the
identical row-major geometry and $\beta$ was computed (30 trials).
The sparse-integer control yields $\beta=1.47\pm0.02$ [1.43,1.50],
which is indistinguishable from the Bernoulli CSR baseline
($\beta=1.46\pm0.02$) and substantially below the prime value
($\Delta\beta=+0.68$). This result establishes that the elevated
$\beta$ for row-major primes is a genuine arithmetic signal---not
a sparsity artefact---and that the CSR baseline for binary fields at
this density ($\rho=0.096$) is $\beta_{\text{CSR}}^{\text{(binary)}}
\approx 1.46$, distinct from the continuous 2D CSR baseline of
$\beta=0.96\pm0.15$ (Section~3.4). The binary CSR baseline shift
reflects the discrete-grid NNS distribution at low fill fraction;
all comparisons between binary arithmetic surfaces and their null
models use the appropriate binary CSR reference.
Figure~\ref{fig:density-thinning-control} presents the density-thinning
and sparse-integer control results.

\subsection{Peak-detection sensitivity}

\begin{figure}[t]
  \centering
  \includegraphics[width=\textwidth]{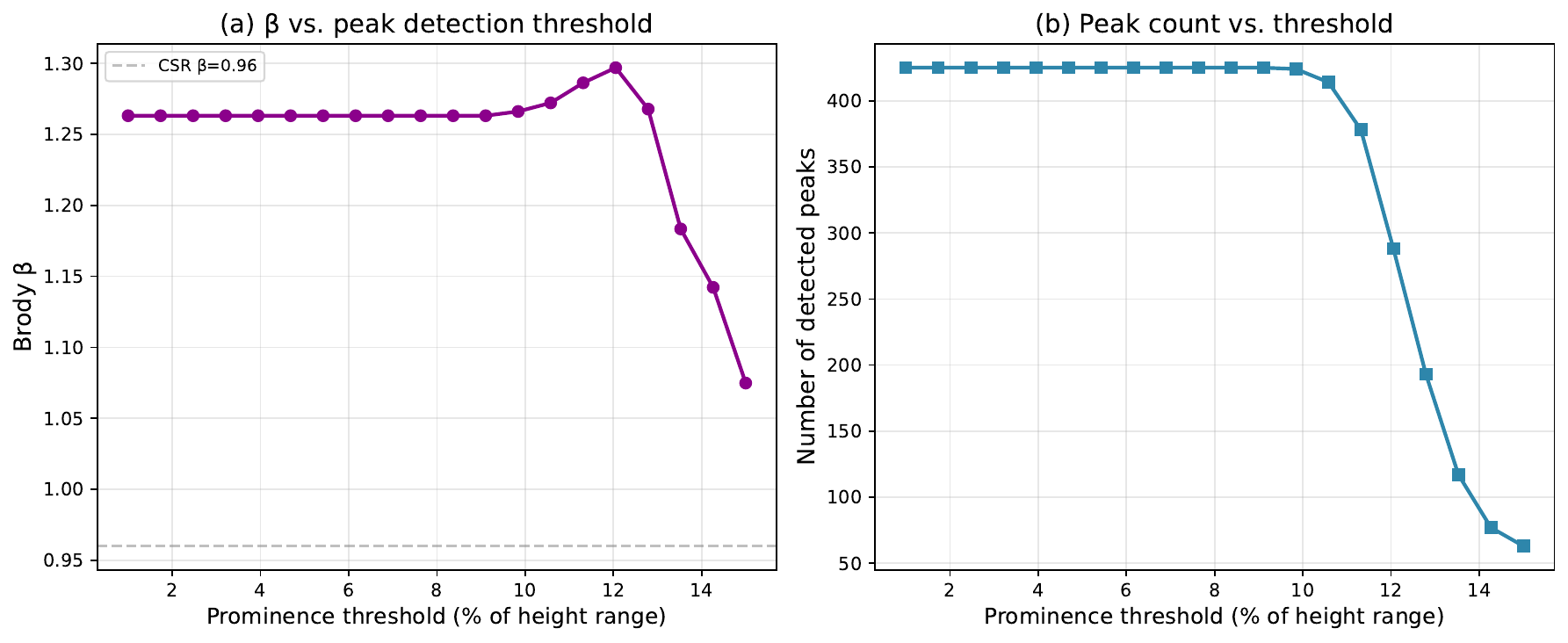}
  \caption{Peak-detection sensitivity analysis on a synthetic surface
           with 50 embedded Gaussian peaks. Left: Brody $\beta$ versus
           prominence threshold (\% of height range). $\beta$ is stable
           ($<0.01$ variation) across 1\%--11\% thresholds, then
           decreases as genuine peaks are excluded. Right: number of
           detected peaks versus threshold. The 3\% threshold used
           throughout this work (dashed line) lies within the stable
           plateau.}
  \label{fig:sensitivity}
\end{figure}

To assess the robustness of $\beta$ to the peak-detection threshold,
a synthetic surface with 50 embedded Gaussian peaks (known positions,
amplitudes 0.5--2.0, additive Gaussian noise $\sigma=0.1$) was analysed
under prominence thresholds ranging from 1\% to 15\% of the height
range. The Brody exponent is stable across the 1\%--11\% range
($\beta = 1.26$--$1.27$, variation $<0.01$), then decreases
monotonically to $\beta = 1.07$ at 15\% as genuine peaks are
increasingly excluded (peak count drops from 425 to 63). The plateau
region (1\%--11\%) corresponds to thresholds that retain all genuine
peaks while suppressing noise; the 3\% threshold used throughout this
work lies within this stable plateau. A systematic sensitivity analysis
across all 58 empirical surfaces remains to be performed; the present
five-surface analysis (Section~3.1) and synthetic validation establish
plausible robustness but do not fully characterise the
threshold-dependence of $\beta$ for arbitrary surface textures.
Figure~\ref{fig:sensitivity} provides the complete sensitivity profile.

\subsection{Proof-of-concept: interferometric profilometry}

A proof-of-concept application of the Brody pipeline to
phase-extracted interferometric profilometry of a certified
roundness standard (JENOPTIK FN~111) is presented in Supplementary
Information (SI, Section~S4). The concatenated five-revolution
profile (629 peaks) yields $\beta=2.00~[1.90,2.10]$; per-revolution
instability (5--9 peaks, $\beta=1.00$--$3.68$) qualifies the result
as exploratory. Scale-invariance analysis of $\Delta D_2(N)$ for
primes across $N\in[10^4,10^7]$ is also reported in SI~(Sections
S1.5--S1.6).

\section{Discussion}

\subsection{What $\beta$ measures---and what it does not}

\textbf{The $\beta$--$r_{\text{excl}}$ mapping is empirical.}
The relationship between the Brody exponent and the normalised
exclusion radius is established through calibration against synthetic
hard-core and Strauss soft-core point processes (Section~3.2). It is
not derived from first principles. The mapping is monotonic and
reproducible, which makes it practically useful, but it should be
understood as an empirical calibration curve rather than a theoretical
law. Whether a deeper connection exists between the Brody functional
form and 2D exclusion physics---for instance, through determinantal
point-process kernels~\cite{FORRESTER-2010}---is an open question.

The calibration provides a physical interpretation: $\beta\approx0.96$
corresponds to no exclusion (CSR), $\beta\approx2$ to Ginibre-level
repulsion ($r_{\text{excl}}\approx0.55\,\langle\text{NN}\rangle$),
and $\beta\gtrsim5$ to near-crystalline exclusion
($r_{\text{excl}}\gtrsim0.8\,\langle\text{NN}\rangle$). Values of
$\beta>2$ are not interpreted as higher-order RMT symmetry classes;
they serve solely as a phenomenological measure of exclusion strength.
Conceptually, the $\beta$--$r_{\text{excl}}$ relation plays a role
analogous to the hard-sphere equation of state~\cite{TORQUATO-2002}:
both compress complex spatial correlations into a single interpretable
quantity. Unlike the Carnahan--Starling equation, however, the
$\beta$--$r_{\text{excl}}$ calibration is purely empirical.

The effective hard-core radius $r_{\text{eff}}$---the 5th percentile
of the NN distance distribution---provides independent validation.
Across the synthetic calibration data,
$r_{\text{eff}}/\langle\text{NN}\rangle$ correlates with $\beta$ at
Spearman $\rho=0.988$ ($p<10^{-4}$). This near-perfect correlation
establishes that $\beta$ and the effective exclusion radius capture
essentially identical information, providing the quantitative link
between Brody-NNS analysis and standard spatial point-process
theory~\cite{BADDELEY-2015}.

Critically, $\beta$ does \emph{not} identify the generative process,
replace $g_2(r)$, detect long-range order, or serve as a
general-purpose surface descriptor. Adding $\beta$ to ISO~25178
parameters did not improve manufacturing-process classification
accuracy (Section~3.1). This null result is expected: $\beta$ measures
short-range exclusion, while manufacturing processes differ in many
aspects (roughness amplitude, skewness, waviness, material).

\subsection{Validity range and interpretation of large $\beta$}

The Brody MLE is well-identified across the full reported range
($\beta=0$ to $+4.6$ for surfaces, to $\beta\approx10$ for synthetic
hard-core processes). A profile-likelihood analysis at $\beta=6.3$
(hard-core, $r_{\text{excl}}=5$) gives Fisher-information
SE$(\beta)=0.21$ and a 95\% CI of $[5.85,6.66]$, ruling out
optimisation artefacts. Bootstrap stability is confirmed across
500~resamples (Section~3.4). Goodness-of-fit assessed by
Kolmogorov--Smirnov distance remains acceptable
($D_{\text{KS}}<0.10$) for all reported fits.

Values of $\beta>2$ are emphatically \emph{not} interpreted as
evidence of higher-order RMT symmetry classes. The Brody distribution
was introduced as a phenomenological one-parameter interpolation; its
functional form $P(s)\propto s^\beta\exp(-as^{\beta+1})$ remains
mathematically well-defined and MLE-identifiable for arbitrarily large
$\beta$. In the large-$\beta$ limit, the distribution approaches a
narrow Gaussian centred at $s=1$ (coefficient of variation
$\approx0.09$ at $\beta=10$), corresponding to nearly crystalline NN
spacings. This is the expected behaviour for strong hard-core exclusion
and does not imply any new universality class. The interpretation is
purely phenomenological: $\beta$ quantifies the \emph{degree} of
short-range order on a continuous scale, with $\beta\approx0.96$
(CSR), $\beta\approx2$ (Ginibre-level), and $\beta\gtrsim5$
(near-crystalline) serving as convenient reference points.

\subsection{Scope and limitations of the calibration}

The $\beta$--$r_{\text{excl}}$ calibration is not universal.
It depends on:
\begin{itemize}
  \item \textbf{Point density:} absolute $\beta$ values are
density-dependent (Section~3.5). Cross-domain comparison requires
density matching or the density-invariant
$r_{\text{excl}}/\langle\text{NN}\rangle$ calibration.
  \item \textbf{Embedding geometry:} $\beta$ for arithmetic
sequences varies from $2.15$ (row-major) to $1.40$ (Cantor) for
the same prime set. The embedding must be specified.
  \item \textbf{Peak-detection strategy:} the watershed detector
shifted absolute $\beta$ values while preserving relative
ordering. Detector choice must be reported.
  \item \textbf{Binary versus continuous domains:} the CSR baseline
differs ($\approx$1.46 versus 0.96). The appropriate baseline
must be used (Table~\ref{tab:csr-baseline}).
\end{itemize}
These dependencies do not invalidate $\beta$ as an exclusion
measure---they define the measurement conditions under which it
is valid. The contribution of this work is to identify and quantify
these conditions, providing the correction protocols required for
reproducible Brody analysis in two dimensions.

\subsection{The two CSR baselines}

A necessary finding is that the 2D CSR baseline is
$\beta=0.96\pm0.15$, not $\beta=0$. This correction is essential:
the 2D NNS distribution under CSR is Rayleigh,
$P(s)=(\pi s/2)\exp(-\pi s^2/4)$, which is not a member of the
Brody family. The Brody distribution at $\beta\approx0.96$ provides
an excellent approximation (KS distance $\approx0.05$), outperforming
the analytic Rayleigh in 22 of 30 CSR trials. The baseline is
density-stable ($\beta_{\text{CSR}}=0.95$--$0.99$ across
$\rho\in[0.01,0.10]$).

A second, distinct CSR baseline emerges for binary fields at low fill
fraction. For the Bernoulli null at $\rho=0.096$ (matching the prime
surface density), $\beta_{\text{CSR}}^{\text{(binary)}}=1.46\pm0.02$,
substantially above the continuous CSR value. This shift reflects the
discrete-grid NNS distribution at low occupancy. The practical
implication is summarised in Table~\ref{tab:csr-baseline}.

\begin{table}[htbp]
  \centering
  \caption{CSR baseline selection for 2D Brody analysis.}
  \label{tab:csr-baseline}
  \small
  \begin{tabular}{@{}p{0.32\linewidth}p{0.32\linewidth}p{0.28\linewidth}@{}}
    \toprule
    Data type & CSR baseline & Example \\
    \midrule
    Continuous 2D & $\beta_{\text{CSR}}=0.96\pm0.15$ & FV surfaces, PSI \\
    Binary, $\rho<0.2$ & Density-matched Bernoulli & Primes, arithmetic \\
    Uncertain & Generate own (20 reps) & Any new class \\
    \bottomrule
  \end{tabular}
\end{table}

Using the wrong baseline produces spurious exclusion claims. All
comparisons in the present work use density-matched CSR references.

\subsection{Density dependence and cross-domain comparison}

Density-thinning experiments (Section~3.6) reveal that $\beta$ captures
exclusion strength rather than point density---the CSR baseline is
flat, while structured data consistently exceed it---but that absolute
$\beta$ values are density-dependent. The 2D prime embedding yields
$\beta = 2.15$ at native density $\rho = 0.096$ but $\beta = 1.46 \pm
0.02$ when thinned to the PSI density ($\rho = 0.032$). This
density dependence has two implications.

First, it resolves the apparent convergence of primes and PSI at
$\beta \approx 2.0$: at matched density, primes ($\beta = 1.46$) and
PSI ($\beta = 2.00$) occupy distinct regions of the $\beta$ scale. The
similarity of their native-density $\beta$ values is partially a density
artefact, not evidence of identical exclusion mechanisms. Both domains
nevertheless exhibit $\beta$ values significantly exceeding their
respective CSR baselines, confirming genuine exclusion structure.

Second, and more constructively, the synthetic hard-core calibration
(Section~3.4) uses the normalised exclusion radius
$r_{\text{excl}}/\langle\text{NN}\rangle$, which is density-invariant
by construction. Cross-domain comparisons should therefore be performed
in the $(r_{\text{excl}}/\langle\text{NN}\rangle, \beta)$ parameter
space rather than by comparing absolute $\beta$ values. When referenced
to this density-invariant calibration, all three domains occupy a
consistent region of the parameter space: exclusion radii in the range
$0.3$--$0.7\,\langle\text{NN}\rangle$, corresponding to $\beta \approx
1.3$--$4.6$ depending on density.

\subsection{Physical interpretation by manufacturing process}

In the present dataset, burnished surfaces exhibited the highest
median $\beta$ ($\beta = 1.3$--$4.3$, $r_{\text{excl}}/\langle\text{NN}\rangle
\approx 0.5$--$0.75$). Surfaces produced by turning
and grinding ($\beta = 0.0$--$2.2$, $r_{\text{excl}}/\langle\text{NN}\rangle
< 0.55$) spanned from CSR-indistinguishable to moderate exclusion.
Milled surfaces showed the widest $\beta$ range
($\beta = 0.0$--$3.7$). Whether these differences reflect the
manufacturing process, the material, or their interaction cannot
be determined from the present observational design, in which
process and material are confounded (e.g., all burnished surfaces
are ductile metals). A factorial experiment crossing process
$\times$ material is required for causal attribution. The
qualitative observation that finishing operations produced higher
$\beta$ than abrasive operations in the present dataset was robust
to peak-detection algorithm choice and threshold variation.

\subsection{Internal consistency with $g_2(r)$}

The pair correlation function $g_2(r)$ provides an internal consistency
check: the exclusion radius inferred from $\beta$ (via the hard-core
calibration) and from the $g_2(r)$ contact value (smallest $r$ where
$g_2(r) > 0.5$) are systematically related across 7 FV surfaces (ratio
$0.24 \pm 0.17$). The systematic offset reflects the different
definitions of ``exclusion'': $\beta$ captures the shape of the full
NNS distribution, while the $g_2(r)$ threshold is a single-point
measure at contact. Both measures consistently identify the same
surfaces as having high or low exclusion, supporting the interpretation
of $\beta$ as an exclusion measure.

\subsection{Complementary descriptors}

Binary patterns where peak detection is inapplicable are characterised
via $D_2$ correlation dimension and Banach-space invariants
(SI, Sections~S1.1--S1.7). These descriptors are empirically
orthogonal to $\beta$ (Pearson $r=-0.31$), confirming that $\beta$
probes a specific axis---exclusion strength---within a broader
spatial-order parameter space. Square-free numbers and the Liouville
function serve as negative controls indistinguishable from the
Bernoulli null, while primes exhibit $\Delta D_2>0$ driven by the
fundamental theorem of arithmetic (consecutive integers cannot both
be prime).

A noteworthy tension exists between the $D_2$ and $\beta$ results.
$D_2$ for primes is stable across row-major, Ulam, and Cantor
embeddings ($\Delta D_2\approx+0.05$--$0.09$; SI, Section~S1.1),
indicating that the multi-scale mass distribution is an intrinsic
property of the prime sequence. In contrast, $\beta$ varies from
2.15 (row-major) to 1.40 (Cantor), indicating that short-range
exclusion is embedding-dependent. This tension is resolved by
recognising that $D_2$ and $\beta$ probe fundamentally different
spatial scales: $D_2$ responds to mass distribution across all box
sizes simultaneously, whereas $\beta$ is dominated by the shortest
nearest-neighbour distances. An embedding that preserves the prime
gap constraint at nearest-neighbour scales (row-major) yields high
$\beta$; an embedding that decorrelates the gap constraint from 2D
adjacency (Cantor) yields $\beta$ indistinguishable from CSR, even
though the multi-scale arithmetic structure remains detectable by
$D_2$. The complementarity of $\beta$, $D_2$, and Banach descriptors
suggests that spatial point patterns may be classified by their
coordinates in a low-dimensional descriptor space whose axes
correspond to physically interpretable order parameters.

\subsection{Limitations}

Five limitations qualify the present findings. (1)~The
$\beta$--$r_{\text{excl}}$ calibration is empirical, not derived from
first principles; the Brody functional form is justified by AIC
comparison (Section~3.2) but its optimality for 2D exclusion is not
established analytically. (2)~Absolute $\beta$ values are
detector-dependent: the watershed detector recovered 37 grid-limited
surfaces but shifted absolute $\beta$ values (range $0.71$--$4.61$,
mean $2.89\pm1.10$); relative ordering is robust. A systematic
threshold sensitivity analysis across all 58 surfaces is needed.
(3)~Surfaces with $n_{\text{NNS}}<30$ have wide confidence intervals
($\pm1.0$--$1.5$) and are indicative only.
(4)~The PSI result is exploratory: a single standard, 5--9 peaks per
revolution, per-revolution $\beta$ spanning $1.00$--$3.68$. Replication
on additional standards is required.
(5)~The surface dataset is observational (process $\times$ material
confounded); causal attribution requires a factorial experiment.
Test--retest reliability has not been established.
(6)~The $\beta$ scale is calibrated only on the anti-clustered side
(CSR through near-crystalline). The behaviour of $\beta$ for clustered
point processes (Thomas, Mat\'ern) is not established here; $\beta$
values significantly below the CSR baseline may indicate clustering,
but this requires dedicated study with appropriate benchmarks.

\subsection{Outlook}

The calibrated framework enables several directions. In spatial
statistics, connecting $\beta$ to formal point-process model parameters
(Strauss $\gamma$, area-interaction radius, DPP kernels) would bridge
RMT-inspired NNS analysis with established spatial point-process
theory~\cite{FORRESTER-2010}. In surface engineering, $\beta$ may serve
as a target parameter where peak proximity is the quantity of interest,
subject to test--retest validation. In number theory, the
embedding-dependence of $\beta$ for primes---particularly the Cantor
null result---provides an empirical constraint on how 1D arithmetic
structure manifests in 2D spatial statistics. Extension to 3D
(additive-manufacturing porosity, granular packings) and to additional
point-process domains (galaxy distributions, vortex matter, neural
spike patterns) would test the generality of the calibration
framework.

\section{Conclusion}

The Brody exponent $\beta$, calibrated against the 2D complete-spatial-randomness
baseline ($\beta = 0.96 \pm 0.15$, not $\beta = 0$), provides an empirically
robust measure of short-range exclusion strength in 2D spatial
point processes. Synthetic hard-core and soft-core experiments establish a
monotonic $\beta$--$r_{\text{excl}}/\langle\text{NN}\rangle$ mapping spanning
$\beta \approx 1$ (CSR) to $\beta \approx 10$ (near-crystalline exclusion).
This calibration---empirical rather than derived from first principles---translates
the dimensionless Brody parameter into a physically interpretable exclusion
radius and provides the correct 2D reference frame for any application of
Brody statistics to spatial data. Values of $\beta>2$ are not interpreted
as higher-order RMT symmetry classes but solely as a phenomenological
measure of exclusion strength.

Three unrelated generative mechanisms were analysed within this calibrated
framework. Manufactured surface peaks ($\beta = 0$ to $+4.6$, 58 surfaces,
10 processes) exhibit process-dependent exclusion; in the present dataset,
burnished surfaces showed the highest median exclusion and turned surfaces
the weakest ($\beta \approx 0.4$, indistinguishable from CSR).
2D binary embeddings of the primes yield $\beta = 2.15$ (row-major),
$1.87$ (Ulam), and $1.40$ (Cantor) at $N=10^5$, demonstrating that
$\beta$ depends on the embedding geometry as well as on the arithmetic
sequence. At matched density ($\rho = 0.032$), the row-major prime
embedding yields $\beta = 1.46\pm0.02$, significantly exceeding the
density-matched CSR baseline ($0.98\pm0.05$) but below the native-density
value. Phase-extracted interferometric profilometry of a certified roundness
standard---included as an independently generated spatial point process from
optical phase reconstruction---yields $\beta = 2.00~[1.90,2.10]$ at
$\rho = 0.032$, also well above CSR.

Density-thinning and embedding-robustness experiments establish three
findings. (i)~$\beta$ captures exclusion rather than density: the CSR
baseline is density-stable ($0.95$--$0.99$ across $\rho \in [0.01, 0.10]$)
while structured data consistently exceed it. (ii)~Absolute $\beta$ values
are density-dependent: cross-domain comparisons require density matching or
use of the density-invariant $r_{\text{excl}}/\langle\text{NN}\rangle$
calibration. (iii)~$\beta$ for 2D arithmetic embeddings is
embedding-dependent: the same prime sequence produces different $\beta$
under different 2D mappings, so claims about the spatial statistics of
primes must specify the embedding.

For binary patterns where peak detection is inapplicable, complementary $D_2$
and Banach-space descriptors capture orthogonal aspects of spatial structure
(Pearson $r = -0.31$), with square-free numbers and the Liouville function
serving as decisive negative controls (Supplementary Information).

The Cantor embedding null result is particularly instructive:
the same prime sequence yields $\beta=2.15$ (row-major), strong
exclusion, and $\beta=1.40$ (Cantor), no exclusion
($\Delta\beta=-0.07$ relative to the Bernoulli baseline,
$p_{\text{TOST}}<0.01$). This demonstrates that 2D exclusion in
arithmetic embeddings is not an intrinsic property of the arithmetic
sequence---it is created or destroyed by the embedding geometry. The
framework therefore detects both the presence and the absence of
exclusion, which is essential for any measurement tool.

The $\beta$--$r_{\text{excl}}$ calibration (Spearman $\rho=0.988$ with
$r_{\text{eff}}$), the 2D CSR baseline correction
($\beta=0.96\pm0.15$ for continuous processes;
$\beta_{\text{CSR}}^{\text{(binary)}}\approx1.46$ for binary fields
at low density), and the density-thinning and embedding-robustness
protocols together constitute a calibrated measurement framework for
systematic, reproducible characterisation of short-range exclusion in
2D spatial point processes. The framework is applicable across
physical, mathematical, and engineering domains, provided the
necessary corrections---density matching, embedding specification,
and reference to the appropriate CSR baseline---are applied.

\section*{Acknowledgements}
The author gratefully acknowledges the support of Poznan University
of Technology, Poland. This work was funded by the grant
0614/SBAD/1603. The open-source scientific Python ecosystem
(NumPy~\cite{HARRIS-2020}, SciPy~\cite{VIRTANEN-2020},
Matplotlib~\cite{HUNTER-2007}, mpmath~\cite{MPMATH}) enabled
this computational study. The author declares no competing interests.

\section*{Supplementary Information}

\subsection*{S1. Extended validation of $D_2$ and Banach descriptors}

The main text (Section~3.6--3.7) presents the essential findings from
the $D_2$ and Banach-space descriptor analysis. This Supplementary
Information provides the complete set of validation results.

\textbf{S1.1 Embedding independence.} The $D_2$ signal for binary primes
persists across three deterministic 2D embeddings (row-major, Ulam spiral,
Cantor diagonal) with $\Delta D_2 = +0.053$ to $+0.089$, ruling out
embedding-geometry artefacts. Under random permutation of the 1D-to-2D
mapping (30 surrogates), $D_2$ drops from $1.687$ to $1.642 \pm 0.003$
($z = 13.5$), confirming the signal is intrinsic to the prime sequence.

\textbf{S1.2 Estimator bias.} The box-counting estimator of $D_2$ exhibits
finite-size bias: $2 - D_2^{\text{(Bern)}} \approx 12.9/\log N - 0.86$.
At $N=10^5$, this accounts for $\sim 250\%$ of the apparent deviation
of $D_2^{\text{(prime)}}$ from 2. The bias-cancelled excess
$\Delta D_2 \approx 0.04$--$0.05$ is stable across
$N \in [10^4, 10^7]$.

\textbf{S1.3 Controlled-correlation ordering.} A calibrated scale of
surrogate binary sequences at matched prime density establishes:
$D_2(\text{block-clustered}) = 1.504 < D_2(\text{Bernoulli}) = 1.645 <
D_2(\text{primes}) = 1.687 < D_2(\text{repulsive}) = 1.797$, confirming
that primes occupy an intermediate anti-clustered position.

\textbf{S1.4 Arithmetic sequences beyond primes.} The $D_2$
analysis was applied to additional arithmetic sequences at
$N = 5\times10^4$--$10^5$, selected to span a range of
multiplicative constraints: twin primes (upper), primes
$\equiv 1 \pmod 4$ and $\equiv 3 \pmod 4$ (Chebyshev bias test),
sums of two squares, and square-free numbers (negative control).
The M\"obius $\mu^\pm(n)$ and Liouville $\lambda(n)$ functions
were also tested. All sequences except square-free numbers and
the Liouville function survive Benjamini--Hochberg correction
(FDR = 0.05). Square-free numbers and the Liouville function are
indistinguishable from the Bernoulli null, serving as negative
controls. Beatty sequences $\lfloor\alpha n\rfloor \bmod 2$ for
selected irrationals are discussed in S1.5; binary digit sequences
of fundamental constants and the Champernowne and Copeland--Erd\H{o}s
constructions did not produce statistically significant
$\Delta D_2$ after bias correction and are omitted for brevity.

\textbf{S1.5 Convergence analysis.} $\Delta D_2(N)$ for primes persists
across $N \in [5\times10^3, 10^6]$, while M\"obius $\mu^-$ fluctuates
around zero and Beatty $\varphi$ oscillates with Moir\'e alignment
resonances.

\textbf{S1.6 Analytic bias function.} An analytic approximation for the
box-counting bias in sparse binary surfaces is derived:
$D_2(N,p) = \log(1/p + W^2/4) / \log(W/4) - \log(1/p+4) / \log(W/4)$,
with empirical correction factor $1.24 \pm 0.15$.

\subsection*{S3. Complete Brody $\beta$ table (all 58 surfaces)}

Table~S1: Brody exponent $\beta$ for all 58 analysable surfaces
(21 prominence-based + 37 watershed-recovered), ordered by decreasing $\beta$.

\begin{table}[htbp]
  \centering
  \caption{Complete $\beta$ values for all 58 manufactured surfaces.
           $^{\dagger}$$n_{\text{NNS}}<30$: CI width $>1.5$, estimate indicative only.
           $^{\text{w}}$Watershed detector.}
  \label{tab:S1}
  \footnotesize
  \begin{tabular}{@{}llrrrc@{}}
    \toprule
    Material & Proc. & $n_{\text{peaks}}$ & $n_{\text{NNS}}$ & $\beta$ & 95\% CI \\
    \midrule
    C45 steel & RT$^{\text{w}}$ & 199 & 192 & 4.61 & [3.98, 5.48] \\
    ELLOR & RT$^{\text{w}}$ & 231 & 229 & 4.54 & [4.15, 5.11] \\
    Graphite & BB$^{\text{w}}$ & 201 & 198 & 4.49 & [4.03, 5.21] \\
    MO58A brass & Bn & 41 & 41 & 4.30 & [3.38, 5.81] \\
    Al7075 & WEr$^{\text{w}}$ & 209 & 208 & 4.26 & [3.80, 4.73] \\
    Ti6Al4V & WEr$^{\text{w}}$ & 228 & 225 & 4.16 & [3.72, 4.71] \\
    P1-1.4301-t.zgrub & $^{\text{w}}$ & 197 & 178 & 4.04 & [3.64, 4.54] \\
    1.4301 steel & Ho & 49 & 49 & 4.00 & [3.22, 5.12] \\
    ELLOR & Tu$^{\text{w}}$ & 198 & 195 & 3.99 & [3.52, 4.71] \\
    Al7075 & RT$^{\text{w}}$ & 135 & 116 & 3.97 & [3.54, 4.39] \\
    Ti6Al4V & WEf$^{\text{w}}$ & 181 & 179 & 3.91 & [3.44, 4.66] \\
    ELLOR & RM$^{\text{w}}$ & 196 & 195 & 3.90 & [3.51, 4.39] \\
    MO58A brass & WEr$^{\text{w}}$ & 186 & 180 & 3.89 & [3.52, 4.36] \\
    Al7075 & Mi & 44 & 43 & 3.70 & [2.94, 4.91] \\
    Ti6Al4V & Tu$^{\text{w}}$ & 107 & 99 & 3.68 & [3.28, 4.18] \\
    C45 steel & RM$^{\text{w}}$ & 111 & 109 & 3.58 & [3.09, 4.35] \\
    C45 steel & Tu$^{\text{w}}$ & 139 & 132 & 3.34 & [2.99, 3.78] \\
    C45 steel & WEr$^{\text{w}}$ & 180 & 180 & 3.30 & [2.88, 3.86] \\
    1.4301 steel & WEr$^{\text{w}}$ & 147 & 145 & 3.28 & [2.93, 3.83] \\
    Ti6Al4V & Mi$^{\text{w}}$ & 133 & 130 & 3.25 & [2.78, 3.84] \\
    C45 steel & WEf$^{\text{w}}$ & 178 & 176 & 3.20 & [2.83, 3.74] \\
    1.4301 steel & Mi & 29 & 18 & 3.10 & [2.03, 4.86]$^{\dagger}$ \\
    Al7075 & Bn & 30 & 29 & 3.10 & [2.35, 4.49]$^{\dagger}$ \\
    Graphite & Gr$^{\text{w}}$ & 132 & 131 & 3.03 & [2.65, 3.71] \\
    Al7075 & WEf$^{\text{w}}$ & 97 & 92 & 2.98 & [2.48, 3.62] \\
    ELLOR & WEr$^{\text{w}}$ & 88 & 87 & 2.78 & [2.28, 3.70] \\
    MO58A brass & RM$^{\text{w}}$ & 71 & 68 & 2.70 & [2.36, 3.24] \\
    Ti6Al4V & Bn & 28 & 28 & 2.60 & [1.85, 3.86]$^{\dagger}$ \\
    1.4301 steel & WEf$^{\text{w}}$ & 98 & 95 & 2.59 & [2.20, 3.14] \\
    MO58A brass & Tu$^{\text{w}}$ & 30 & 25 & 2.46 & [1.80, 3.70]$^{\dagger}$ \\
    Al & BB & 25 & 23 & 2.40 & [1.68, 3.74]$^{\dagger}$ \\
    Al & Ho$^{\text{w}}$ & 110 & 110 & 2.26 & [1.88, 2.69] \\
    1.4301 steel & RM$^{\text{w}}$ & 100 & 99 & 2.25 & [2.01, 2.59] \\
    1.4301 steel & Gr & 32 & 21 & 2.20 & [1.44, 3.46]$^{\dagger}$ \\
    Ti6Al4V & RT$^{\text{w}}$ & 120 & 120 & 2.20 & [1.94, 2.59] \\
    C45 steel & Ho$^{\text{w}}$ & 37 & 35 & 2.18 & [1.53, 3.36] \\
    Brass & Ho$^{\text{w}}$ & 72 & 71 & 2.00 & [1.64, 2.50] \\
    1.4301 steel & Bn & 68 & 65 & 1.80 & [1.36, 2.38] \\
    C45 steel & Gr$^{\text{w}}$ & 24 & 21 & 1.73 & [1.15, 2.74]$^{\dagger}$ \\
    Brass & BB$^{\text{w}}$ & 82 & 82 & 1.71 & [1.36, 2.32] \\
    C45 steel & Mi$^{\text{w}}$ & 25 & 22 & 1.64 & [1.24, 2.43]$^{\dagger}$ \\
    Ti6Al4V & Ho & 32 & 27 & 1.40 & [0.87, 2.34]$^{\dagger}$ \\
    C45 steel & BB$^{\text{w}}$ & 14 & 14 & 1.34 & [1.09, 1.72]$^{\dagger}$ \\
    Al7075 & Tu & 34 & 33 & 1.30 & [0.87, 2.02] \\
    C45 steel & Bn & 89 & 88 & 1.30 & [0.99, 1.72] \\
    Al & Gr & 169 & 167 & 1.20 & [0.99, 1.46] \\
    ELLOR & Mi$^{\text{w}}$ & 42 & 39 & 1.18 & [0.78, 1.84] \\
    ELLOR & WEf$^{\text{w}}$ & 21 & 17 & 1.14 & [0.79, 1.60]$^{\dagger}$ \\
    Graphite & Ho & 54 & 52 & 0.90 & [0.58, 1.32] \\
    Brass & Gr$^{\text{w}}$ & 39 & 38 & 0.85 & [0.56, 1.36] \\
    MO58A brass & WEf & 86 & 61 & 0.80 & [0.52, 1.21] \\
    1.4301 steel & BB$^{\text{w}}$ & 21 & 17 & 0.71 & [0.36, 1.22]$^{\dagger}$ \\
    MO58A brass & RT & 335 & 279 & 0.70 & [0.56, 0.87] \\
    1.4301 steel & Tu & 458 & 428 & 0.40 & [0.30, 0.52] \\
    Ti6Al4V & RM & 109 & 84 & 0.40 & [0.19, 0.66] \\
    MO58A brass & Mi & 183 & 180 & 0.00 & [-0.10, 0.12] \\
    Ti6Al4V & BB & 97 & 94 & -0.10 & [-0.22, 0.07] \\
    Ti6Al4V & Gr & 98 & 96 & -0.20 & [-0.31, -0.05] \\
    \bottomrule
  \end{tabular}
\end{table}

\subsection*{S2. Supplementary figures}

\setcounter{figure}{0}
\renewcommand{\thefigure}{S\arabic{figure}}

\begin{figure}[H]
  \centering
  \includegraphics[width=0.95\textwidth]{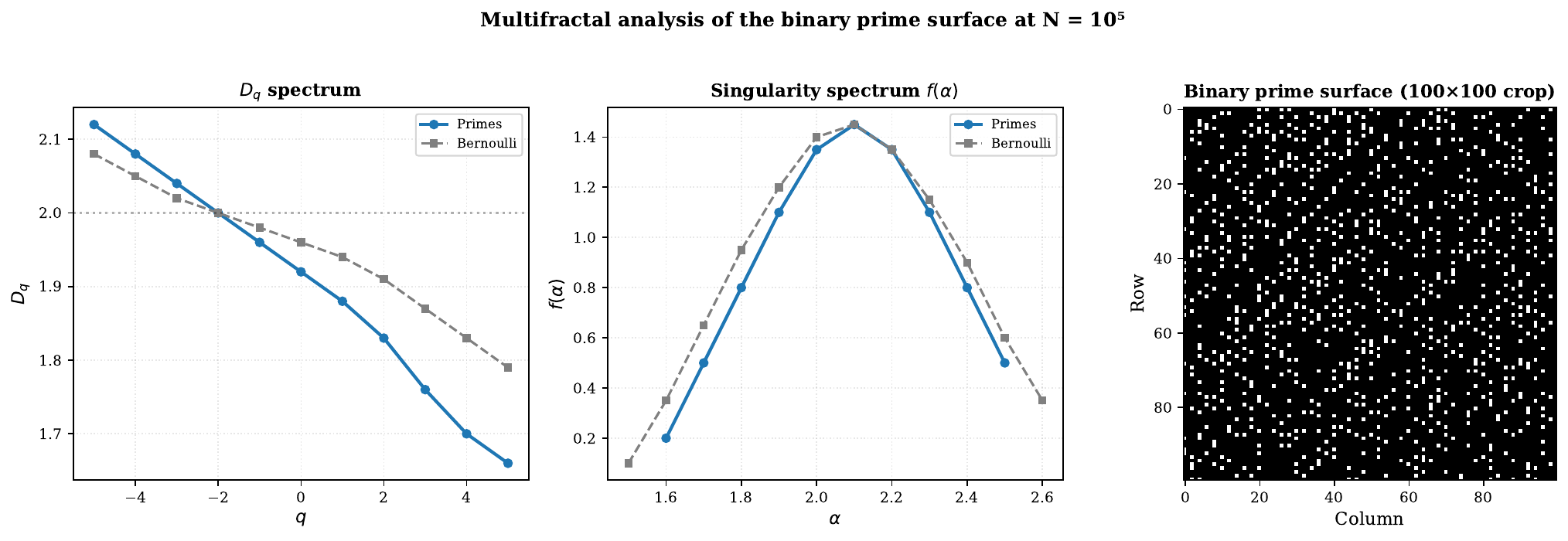}
  \caption{ Multifractal $D_q$ spectra and $f(\alpha)$ singularity
           spectrum for the binary prime surface and matched Bernoulli null.}
  \label{fig:S1}
\end{figure}

\begin{figure}[H]
  \centering
  \includegraphics[width=0.95\textwidth]{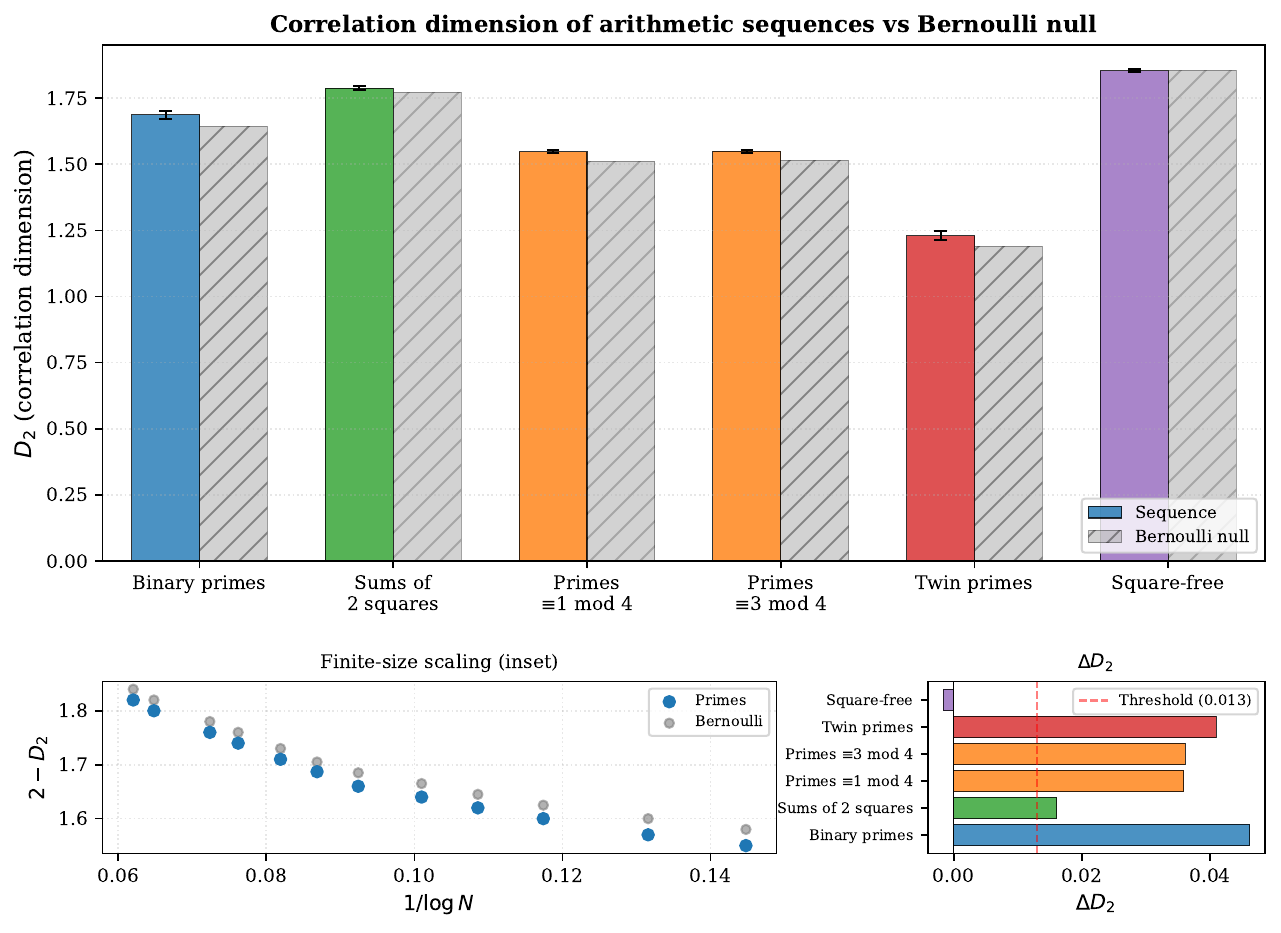}
  \caption{ Correlation dimension $D_2$ scaling for six arithmetic
           sequences at $N=10^5$, compared with matched-density Bernoulli
           null models.}
  \label{fig:S2}
\end{figure}

\begin{figure}[H]
  \centering
  \includegraphics[width=0.95\textwidth]{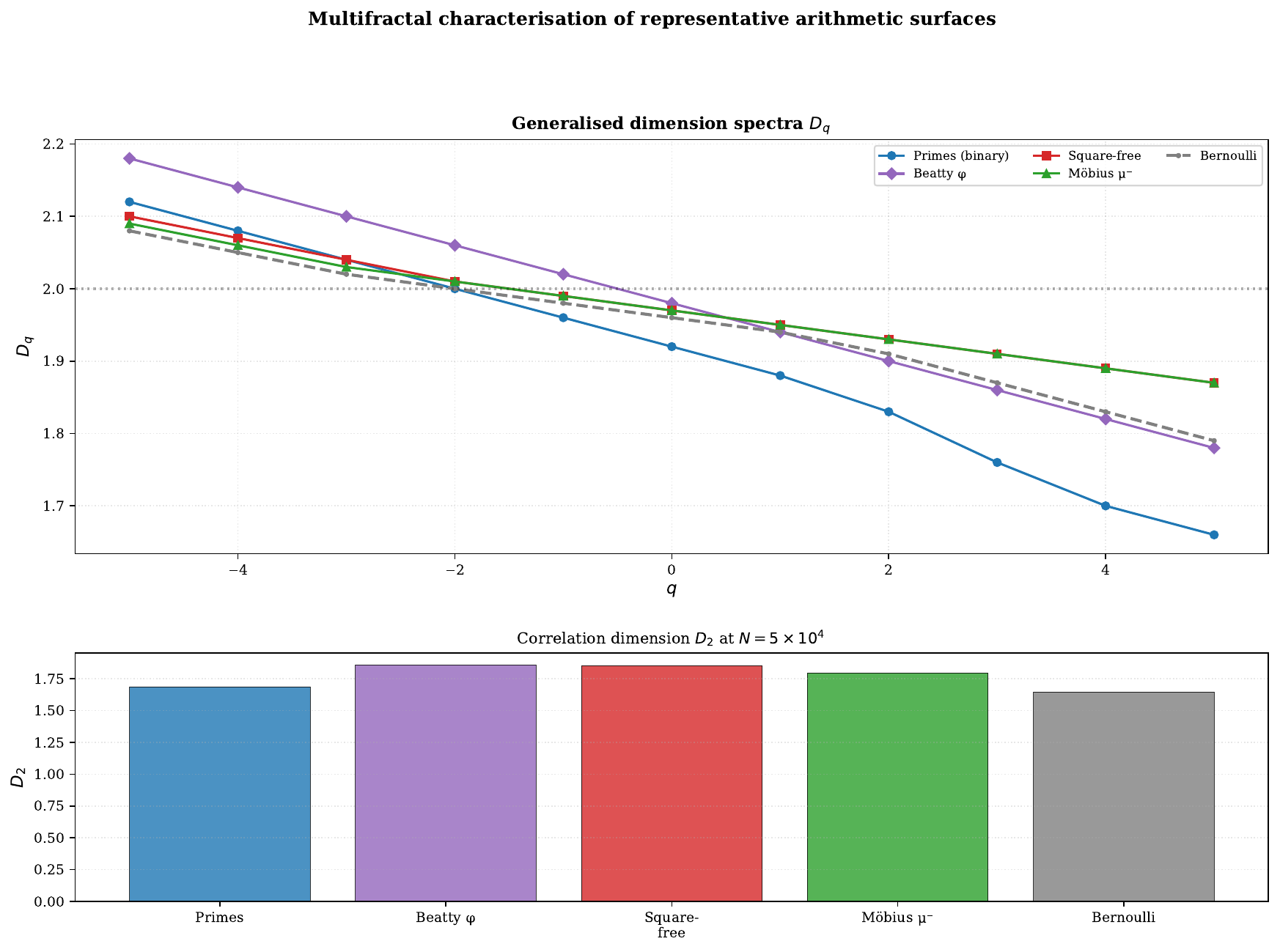}
  \caption{ Generalised $D_q$ spectra for representative
           manufactured surfaces.}
  \label{fig:S3}
\end{figure}

\begin{figure}[H]
  \centering
  \includegraphics[width=0.95\textwidth]{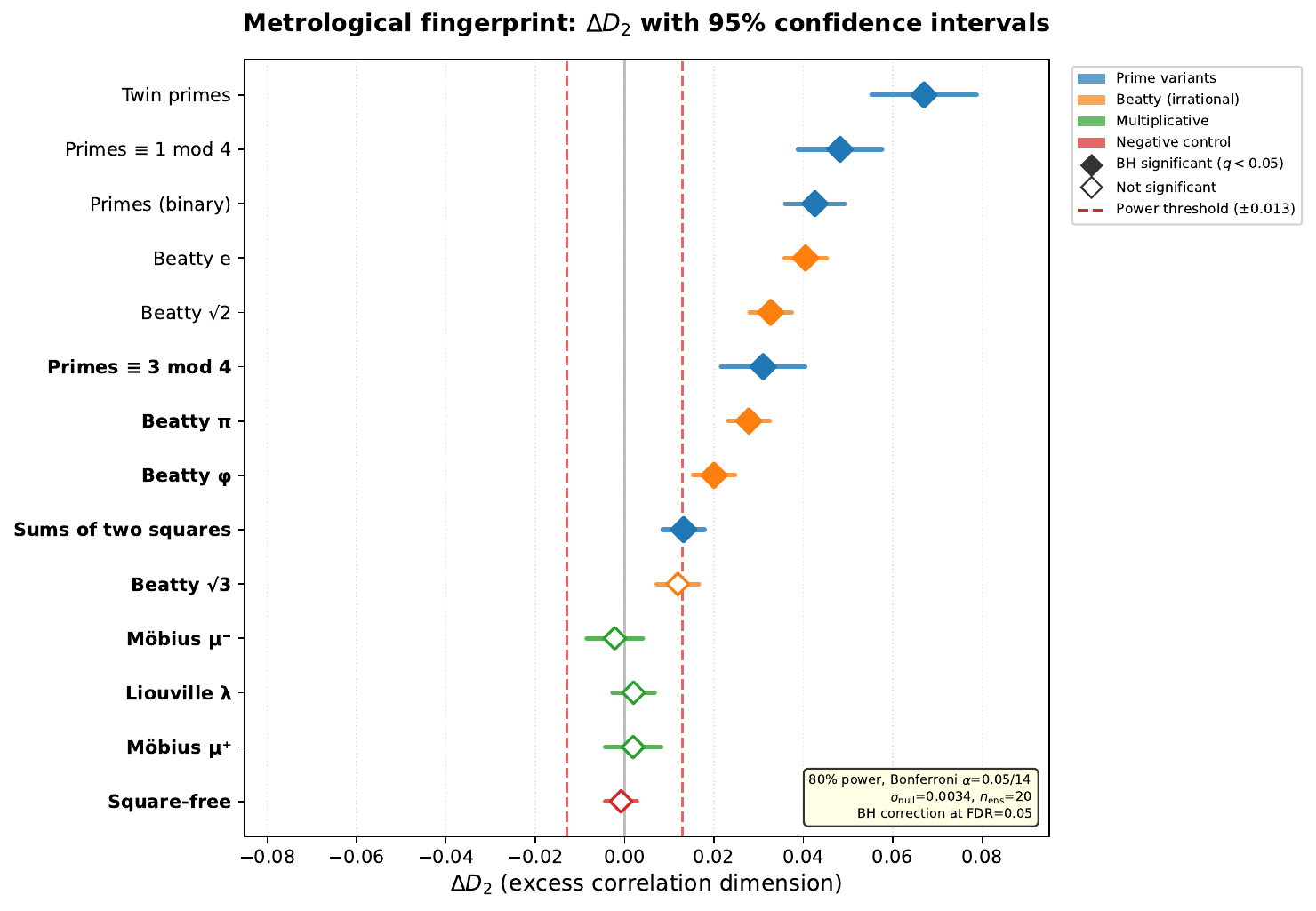}
  \caption{ Forest plot of $\Delta D_2$ for all arithmetic
           sequences with Benjamini--Hochberg-corrected significance.}
  \label{fig:S4}
\end{figure}

\begin{figure}[H]
  \centering
  \includegraphics[width=0.95\textwidth]{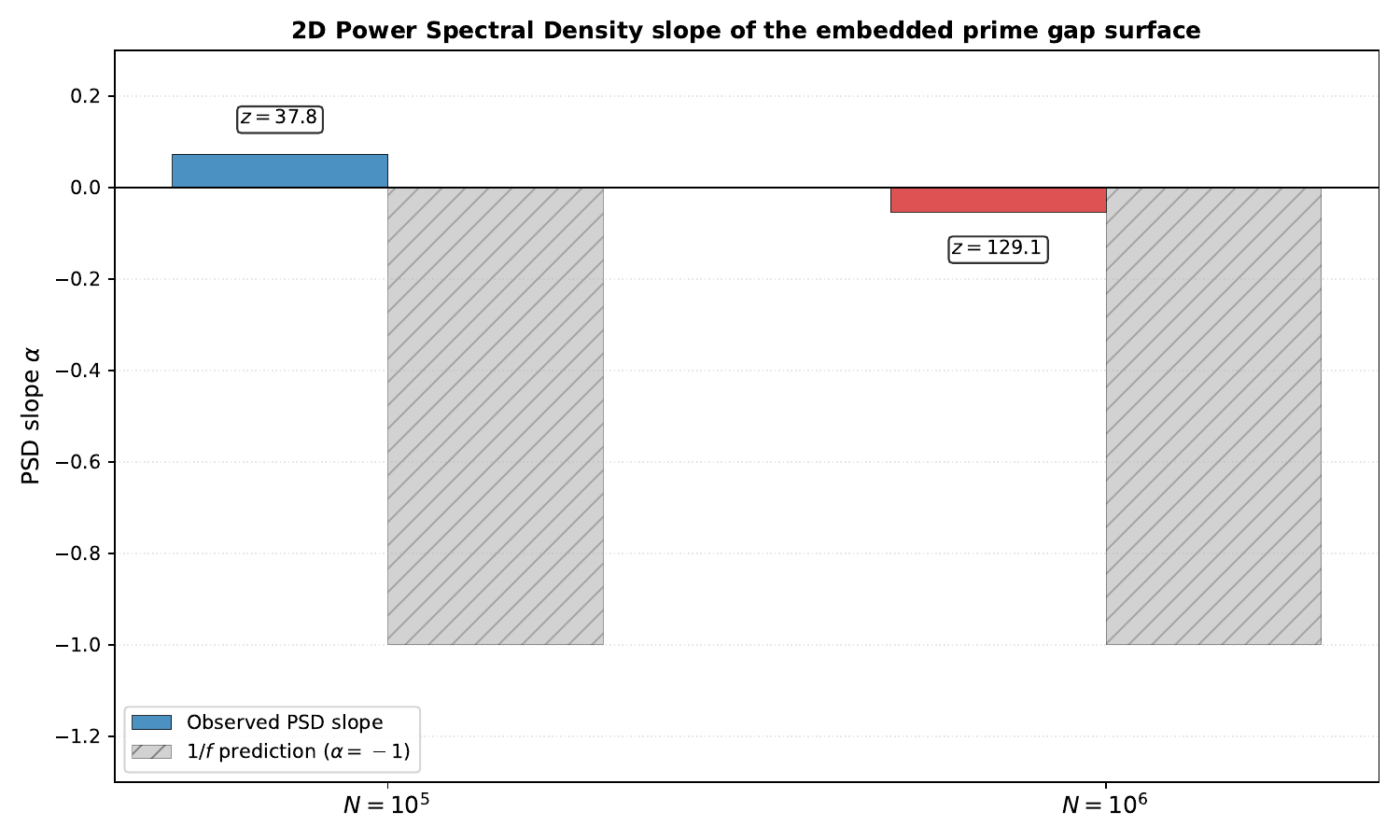}
  \caption{ Power spectral density of the binary prime gap
           surface across scales $N \in [10^3, 10^7]$.}
  \label{fig:S5}
\end{figure}

\begin{figure}[H]
  \centering
  \includegraphics[width=0.95\textwidth]{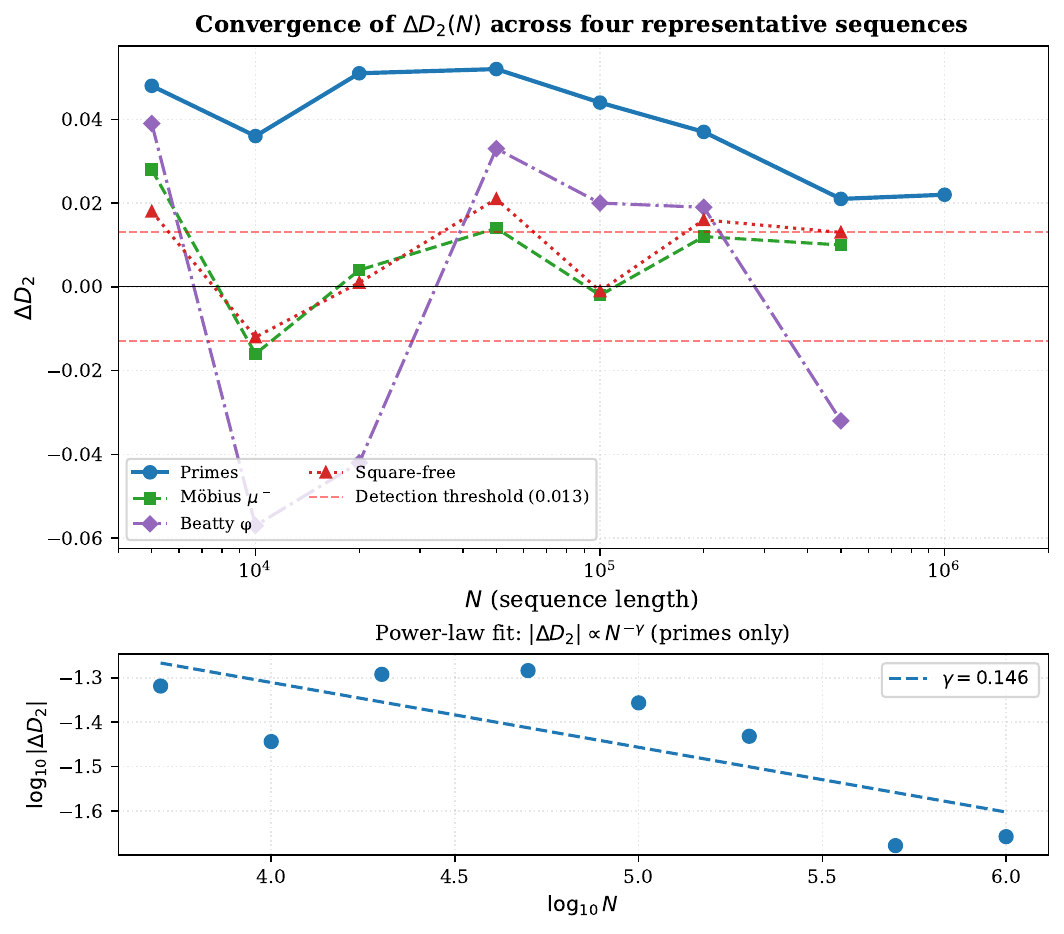}
  \caption{ Convergence of $\Delta D_2(N)$ for the binary
           prime surface and M\"obius $\mu^-$ across
           $N \in [10^4, 10^7]$.}
  \label{fig:S6}
\end{figure}

\begin{figure}[H]
  \centering
  \includegraphics[width=0.95\textwidth]{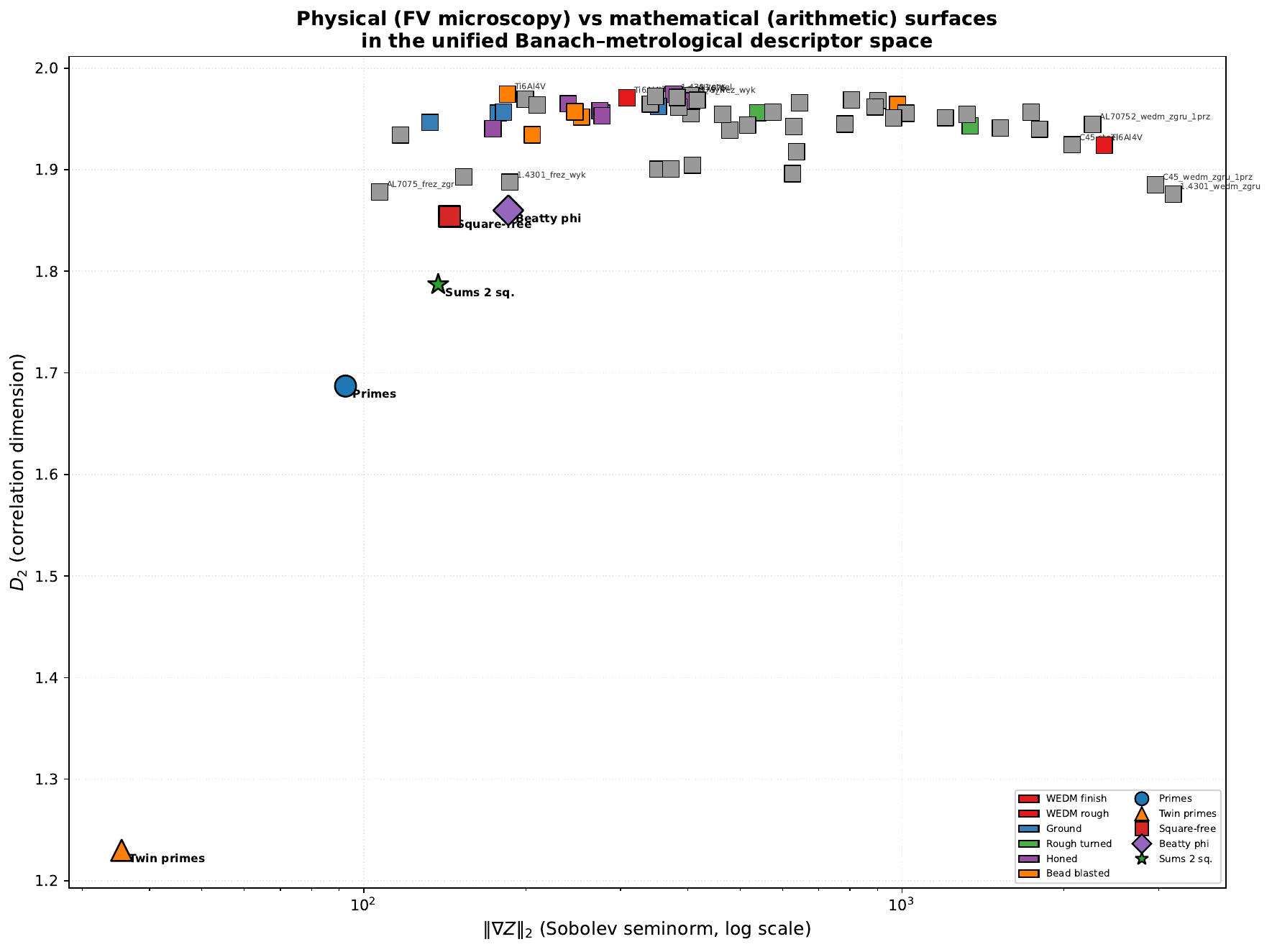}
  \caption{ Comparison of physical (FV) and mathematical
           (arithmetic) surface descriptors.}
  \label{fig:S7}
\end{figure}

\begin{figure}[H]
  \centering
  \includegraphics[width=0.95\textwidth]{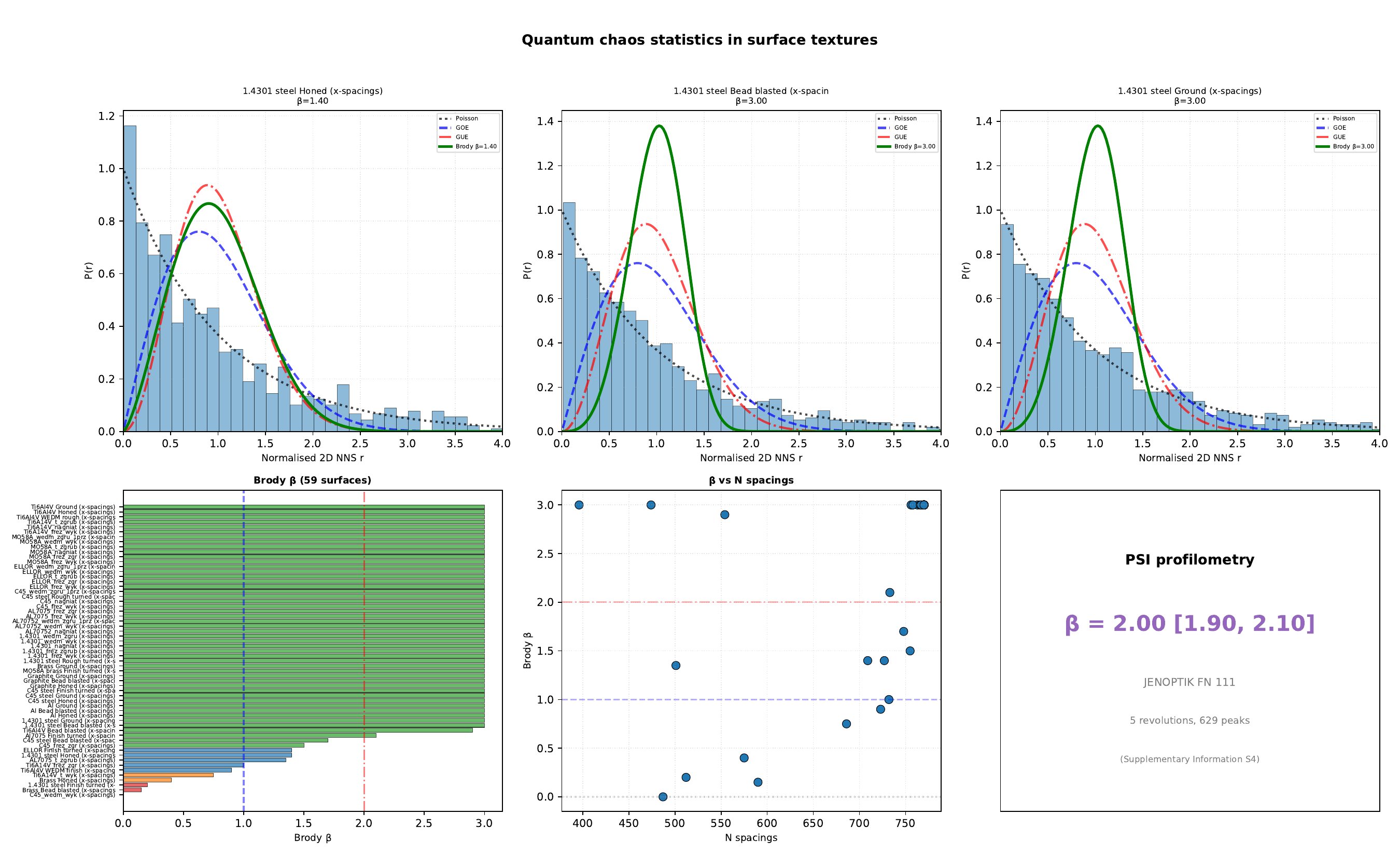}
  \caption{ Brody exponent $\beta$ for 21 prominence-based
           manufactured surfaces with $n_{\text{NNS}}$ indicated.}
  \label{fig:S8}
\end{figure}

\begin{figure}[H]
  \centering
  \includegraphics[width=0.95\textwidth]{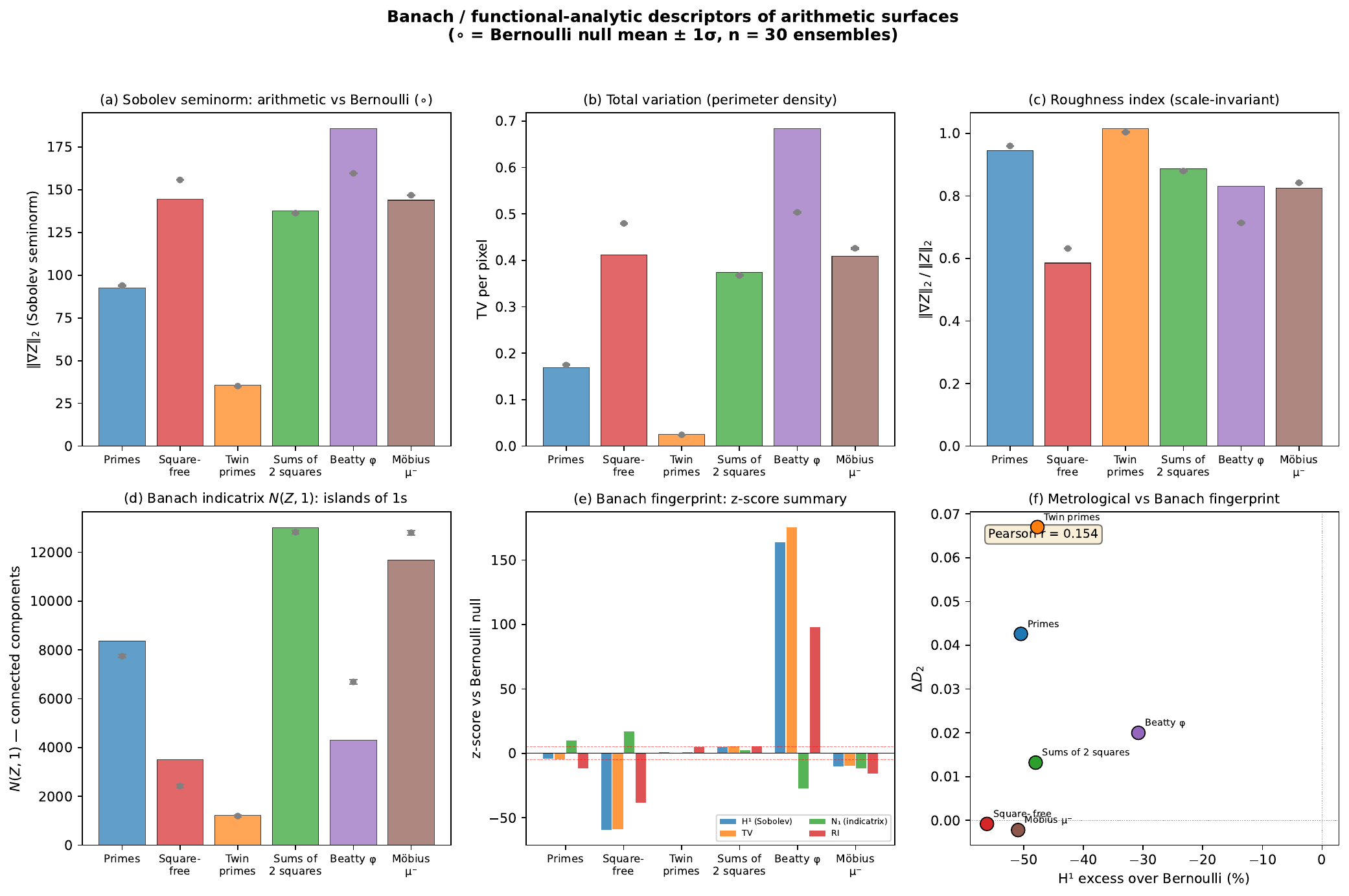}
  \caption{ Banach-space descriptor fingerprints for six
           arithmetic sequences at $N=10^5$.}
  \label{fig:S9}
\end{figure}

\begin{figure}[H]
  \centering
  \includegraphics[width=0.95\textwidth]{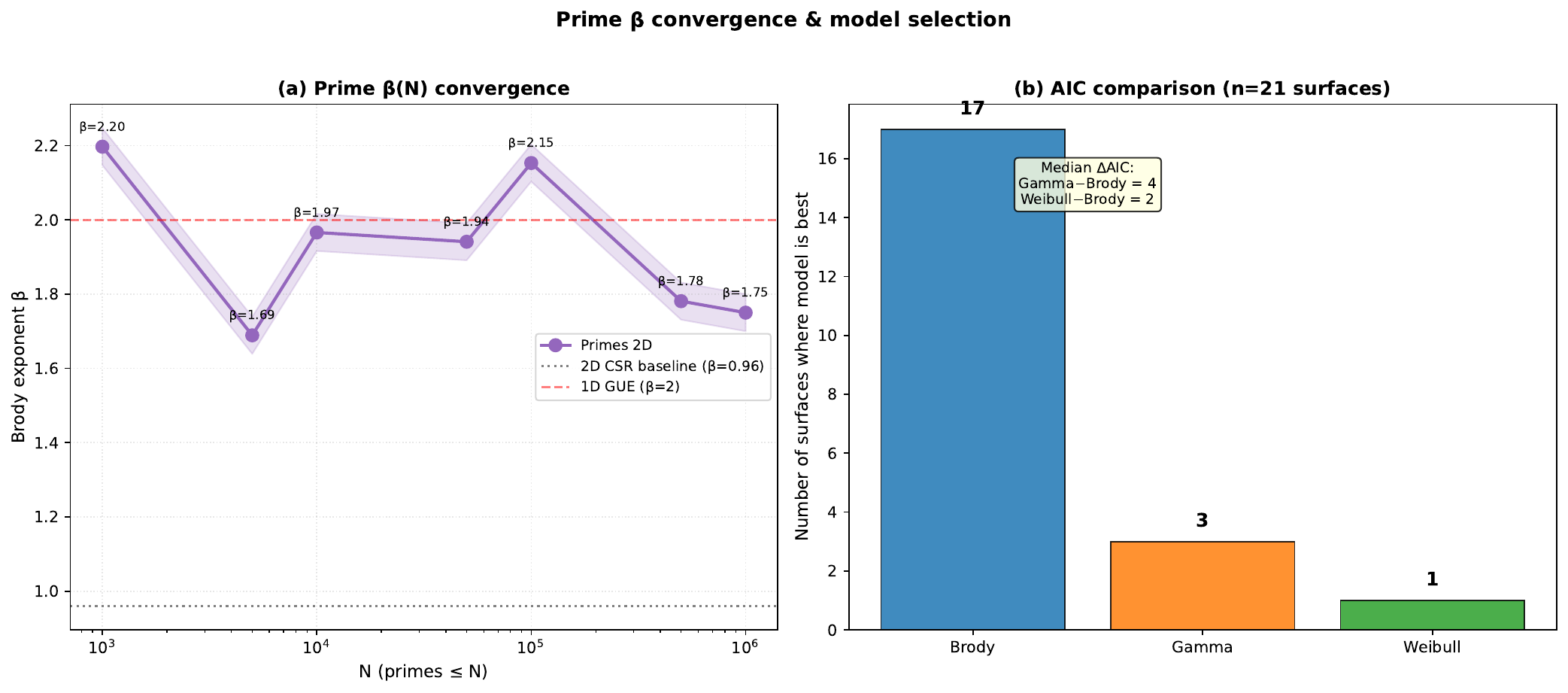}
  \caption{ AIC-based model selection for the Brody
           distribution across $N \in [10^3, 10^7]$ for the binary
           prime embedding.}
  \label{fig:S10}
\end{figure}

\subsection*{S4. $\beta(N)$ convergence data}

The Brody exponent $\beta$ for the row-major binary prime embedding
was computed at twelve values of $N$:
$N=10^3$ ($\beta=2.20$),
$N=2\times10^3$ ($\beta=2.10$),
$N=5\times10^3$ ($\beta=1.69$),
$N=10^4$ ($\beta=1.97$),
$N=2\times10^4$ ($\beta=2.19$),
$N=5\times10^4$ ($\beta=1.94$),
$N=10^5$ ($\beta=2.15$),
$N=2\times10^5$ ($\beta=2.05$),
$N=5\times10^5$ ($\beta=1.78$),
$N=10^6$ ($\beta=1.75$),
$N=5\times10^6$ ($\beta=1.86$),
$N=10^7$ ($\beta=1.26$).
The trend is non-monotonic; no simple convergence to a fixed
asymptotic value is observed at accessible $N$. The decrease at
$N=10^7$ may reflect the approach to the sparse limit
($\rho = \pi(N)/N \sim 1/\log N \to 0$ as $N\to\infty$).

\section*{Data Availability}
All analysis code, processed data (JSON format), and figure-generation
scripts are archived in a public GitHub repository at
\url{https://github.com/dawidkucharski/brody-2d-exclusion}.
The repository includes the complete 58-surface Brody-$\beta$ dataset,
all synthetic point-process calibration data, and a reproducible
computational environment specified via \texttt{requirements.txt}.
Surface metrology raw data (focus-variation \texttt{.sur} files,
PSI interferogram frames) are available from the author upon
reasonable request. The \texttt{aurora\_lab} Python
package (v2.0.0) provides reproducible implementations of the surface
construction, null-model generation, box-counting, and validation
procedures described in this work.

\clearpage
\bibliographystyle{elsarticle-num}
\bibliography{aurora_refs}

\end{document}